\documentclass{aa}  
\usepackage{graphicx}
\usepackage{txfonts}
\usepackage[T1]{fontenc}
\usepackage{ae,aecompl}
\usepackage[flushleft]{threeparttable}
\usepackage{multirow,bigdelim}
\usepackage{mathtools}
\usepackage{siunitx}
\usepackage[colorlinks=true,linkcolor=blue, citecolor=blue]{hyperref}%

\usepackage{subfig}

\usepackage{graphicx}	
\usepackage{amsmath}	
\usepackage{amssymb}	
\usepackage{bm}		
\usepackage{pdflscape}	

\usepackage[T1]{fontenc}
\usepackage{ae,aecompl}

\usepackage{bigints}    
\usepackage{units}
\usepackage{tabularx}
\usepackage{cleveref}
\crefformat{section}{\S#2#1#3} 
\crefformat{subsection}{\S#2#1#3}
\crefformat{subsubsection}{\S#2#1#3}

\usepackage{ulem}
\usepackage[dvipsnames]{xcolor}

\usepackage{bm}
\usepackage{soul}
\usepackage{comment}

\usepackage{color}

\usepackage{etoolbox}
\makeatletter
\patchcmd\@combinedblfloats{\box\@outputbox}{\unvbox\@outputbox}{}{%
  \errmessage{\noexpand\@combinedblfloats could not be patched}%
}%
\makeatother

\newcommand{\be}{\begin{equation}}
\newcommand{\ee}{\end{equation}}
\newcommand{\ba}{\begin{eqnarray}}
\newcommand{\ea}{\end{eqnarray}}

\newcommand{\dk}{\boldsymbol{w}_{g}^{(k)}}

\newcommand{\dbar}{\overline{\boldsymbol{w}}_{g}}
\newcommand{\njk}{N_{\rm JK}}
\newcommand{\healpix}{\mathtt{HEALPIX}}

\begin{document}

\title{Clustering of red-sequence galaxies in the fourth data release of the Kilo-Degree Survey}
\titlerunning{KiDS DR4 LRG clustering}

\author{
Mohammadjavad Vakili\inst{1}\thanks{\emph{E-mail:} vakili@strw.leidenuniv.nl}, Henk Hoekstra\inst{1}, Maciej Bilicki\inst{2}, Maria-Cristina Fortuna\inst{1}, Konrad Kuijken\inst{1}, Angus H. Wright\inst{3}, Marika Asgari\inst{4}, Michael Brown\inst{5}, Elisabeth Dombrovskij\inst{1}, Thomas Erben\inst{6}, Benjamin Giblin\inst{4}, Catherine Heymans\inst{3,4}, Hendrik Hildebrandt\inst{3}, Harry Johnston\inst{7} Shahab Joudaki\inst{8}, Arun Kannawadi\inst{9}.}

\authorrunning{M. Vakili et al.}

\institute{Leiden Observatory, Leiden University, PO Box 9513, Leiden, NL-2300 RA, the Netherlands\and
Center for Theoretical Physics, Polish Academy of Sciences, al. Lotników 32/46, 02-668, Warsaw, Poland\and
Ruhr-Universität Bochum, Astronomisches Institut, German Centre for Cosmological Lensing, Universitätsstr. 150, 44801 Bochum, Germany\and
Institute for Astronomy, University of Edinburgh, Royal Observatory, Blackford Hill, Edinburgh, EH9 3HJ, U.K.\and
School of Physics, Monash University, Clayton, VIC 3800, Australia\and
Argelander-Institut für Astronomie, Universität Bonn, Auf dem Hügel 71, 53121 Bonn, Germany\and
Department of Physics and Astronomy, University College London, Gower Street, London WC1E 6BT, UK\and
Department of Physics, University of Oxford, Denys Wilkinson Building, Keble Road, Oxford OX1 3RH, U.K.\and
Department of Astrophysical Sciences, Princeton University, 4 Ivy Lane, Princeton, NJ 08544, USA
}

\date{Accepted XXX. Received YYY; in original form ZZZ}

\label{firstpage}
\makeatletter
\renewcommand*\aa@pageof{, page \thepage{} of \pageref*{LastPage}}
\makeatother


\abstract{We present a sample of luminous red-sequence galaxies to study the large-scale structure in the fourth data release of the Kilo-Degree Survey. The selected galaxies are defined by a red-sequence template, in the form of a data-driven model of the colour-magnitude relation conditioned on redshift. In this work, the red-sequence template is built using the broad-band optical+near infrared photometry of KiDS-VIKING and the overlapping spectroscopic data sets. The selection process involves estimating the red-sequence redshifts, assessing the purity of the sample, and estimating the underlying redshift distributions of redshift bins. After performing the selection, we mitigate the impact of survey properties on the observed number density of galaxies by assigning photometric weights to the galaxies. We measure the angular two-point correlation function of the red galaxies in four redshift bins, and constrain the large scale bias of our red-sequence sample assuming a fixed $\Lambda$CDM cosmology. We find consistent linear biases for two luminosity-threshold samples (`dense' and `luminous'). We find that our constraints are well characterized by the passive evolution model.}


\maketitle

\section{Introduction}

The Kilo-Degree Survey (KiDS) is an optical galaxy survey primarily designed to map the large-scale structure by studying the weak gravitational lensing of galaxies \citep{kids,kuijken2015, kuijken2019}. This is done by measuring the distortion of the shapes of distant galaxies known as cosmic shear, which has become a cornerstone of modern cosmological imaging surveys. Current surveys are already yielding competitive constraints on some cosmological parameters \citep[e.g.][]{troxel2017, hikage2019, hendrik2020, asgari2020}. 

However, the full constraining potential of weak lensing studies can only be realized through the joint analysis of the cosmic shear of background galaxies and the positions of foreground galaxies with robust distance estimates. This involves measuring the correlation between the cosmic shear estimates of the background galaxies, the correlation between the positions of foreground galaxies, as well as the cross-correlation between the cosmic shear of background galaxies and the positions of foreground galaxies, known as `galaxy-galaxy lensing' \citep[e.g.][]{cacciato2013, des_y1_cosmology, edo2016, joudaki2018}. 
Such combined analyses yield tighter constraints on cosmological parameters, and offer a venue for mitigation of a range of observational and theoretical systematics such as photometric redshift uncertainties and intrinsic alignments (\citealt{edo2016, joudaki2018, sam2019, heymans2020}). 

In this work, we focus on selecting a sample of galaxies with robust redshift estimates as well as measuring their angular two-point correlation function in slices of redshift. Following \citet{vakili2019} we construct a sample of red-sequence galaxies by leveraging the fact that the distribution of these galaxies in colour space closely follows a multivariate Gaussian distribution. The mean of this distribution is a linear function of magnitude. Furthermore, the coefficients of this linear relation, as well as the covariance of the Gaussian distribution, are determined by the redshift \citep[e.g.][]{bower1992,ellis1997,gladders1998,stanford1998}. 

We can then leverage this empirical distribution to select red-sequence galaxies with the broad-band photometry of imaging surveys (\citealt{gladders_yee2000,hao2009,redmap_sdss,rozo2016,elvin2017,oguri2018,vakili2019}). In this work, we build this data-driven model with the multi-band photometry of the KiDS Data Release 4 (DR4, \citealt{kuijken2019}) and its overlap with the following spectroscopic data sets: SDSS DR13 (\citealt{sdss_dr13}), GAMA (\citealt{driver2011}), 2dFLenS (\citealt{blake2016}), and the GAMA reanalysis of the redshifts in the COSMOS region (hereafter G10-COSMOS, \citealt{davis2015}). 


Following the $\textsc{redMagiC}$ prescription (\citealt{rozo2016}), we impose a set of luminosity ratio cuts and constant comoving densities, and we construct two samples of red-sequence galaxies with nearly constant comoving density suitable for galaxy clustering and galaxy-galaxy lensing studies. We call these the dense (high density, low brightness) and the luminous (low density, high brightness) samples. The former (latter) sample is constructed such that the comoving density is approximately $10^{-3}$ $\mathrm{Mpc}^{-3}h^3$ ($2.5\times10^{-4}$ $\mathrm{Mpc}^{-3}h^3$). The main differences between the red-sequence selection in this work and the previous KiDS DR3 work of \citet{vakili2019} are: the inclusion of the VIKING (\citealt{Edge2013, KV450}) $Z$-band magnitudes in the red-sequence template, and the inclusion of the G10-COSMOS in the spectroscopic calibration of the model, adding more depth and redshift coverage for our red-sequence model. In addition, we apply this method to the fourth data release of KiDS which more than doubles the sky coverage with respect to the KiDS DR3. 

Furthermore, we utilize the VIKING $K_{\rm s}$-band magnitude to investigate the impurity (contamination with stellar like objects) of the selected objects within each luminosity threshold sample.
The redshift reach of each sample is chosen such that the sample remains pure while nearly volume-limited (constant comoving density) below that redshift. Afterwards, we divide the galaxy sample into four redshift bins between 0.15 and 0.8, with the first three redshift bins comprising the galaxies in the dense sample and the last bin consisting of the galaxies in the luminous sample. After modeling the individual redshift distributions of red galaxies in our sample with a Student t-distribution, we compute the underlying redshift distributions of galaxies in the four redshift bins by summing the individual redshift probabilities.

Given that galaxy clustering measures the excess probability of finding pairs of galaxies at a given angular or physical separation, accounting for the impact of survey properties on the galaxy density variations across the footprint requires a careful treatment. These properties 
can influence the detection of galaxies as well as the selection process of any galaxy sample in the survey \citep[e.g.][]{morrison2015,alam2017,kwan2017,ross2017,elvin2017,crocce2019,kalus2019}. 
In order to remove the dependence of the on-sky variations of galaxy number density on the KiDS survey properties, we assign a set of photometric weights for galaxies in each redshift bin separately. By up-weighting (down-weighting) areas of the survey where the galaxy density is down-graded (enhanced) due to survey properties, this scheme mitigates the systematic modes present in the sample. In our method, similar to that of \citet{bautista2018sdss, icaza2020clustering}, the photometric weights are estimated such that possible correlations between the survey properties are accounted for. 

After measuring the angular clustering signal and its covariance in each of the redshift bins, we estimate the large-scale bias of these galaxies assuming a fixed $\Lambda$CDM cosmology. We then compare our bias constraints with the predictions of the passive evolution bias model of \citet{Fry1996}.

In the last step of our analysis, i.e. estimation of the galaxy bias parameters, we use a blinding method in order to prevent the final bias constraints from influencing the choices we have made in our work. Prior to estimating the galaxy bias parameters, we blind our covariance matrices of the clustering signals with the method proposed by \citet{sellentin2019}. The added advantage of this approach is that the intermediate steps of our data analysis pipeline such as catalogue curation and systematic mitigation remain unchanged.

This paper is structured as follows. The data, both photometric and spectroscopic, are described in Sect.~\ref{sec:data}. In Sect.~\ref{sec:selection} we discuss the sample selection and the photometric redshifts. We then provide the galaxy-density systematic correlations and the 
derivation of photometric weights in Sect.~\ref{sec:systematic}. In Sect.~\ref{sec:clustering} we present the angular two-point correlation functions as well as the theoretical predictions. Finally, we summarize and conclude in Sect.~\ref{sec:summary}. 

\begin{table*}
	\centering
	\caption{Summary of the spectroscopic data used in this work.} 
	\label{tab:zspec}
	\begin{tabularx}{1.95\columnwidth}{lcccccccr} 
		\hline
		Data &  \# objects in KiDS & \# unique objects & $z_{16\%}$ & $z_{50\%}$ & $z_{84\%}$ & $m_{r, 16\%}$ & $m_{r, 50\%}$ & $m_{r, 84\%}$\\
		\hline
		GAMA     & 233046 & 233046 & 0.12  & 0.22 & 0.34 & 18.7 & 19.6 & 20.1  \\
		SDSS     & 99253 &  77371  & 0.09  & 0.37 & 0.57 & 17.7 & 19.7 & 21.1  \\
        2dFLenS  & 37462 &  34253  & 0.13  & 0.30 & 0.59 & 18.3 & 19.5 & 21.0  \\
        COSMOS (GAMA-G10)   & 20324 &  20324  & 0.32 & 0.68 & 1.24 & 21.6 & 22.9 & 24.0 \\
		\hline
	 \end{tabularx}
    \tablefoot{The first three columns are the the spectroscopic data sets under consideration, their total number of objects in KiDS DR4, and the unique number of objects in each data set. The next three columns are the 16-, 50-, and 84-percentiles of the spectroscopic redshifts of the unique objects in the spectroscopic data sets. The last three columns are the 16-, 50-, and 84-percentiles of the KiDS GAaP $r$-band magnitudes of the objects in each spec-$z$ data. For each row, the unique number of objects is obtained after removing the objects that are already included in the GAMA catalogue (GAMA or SDSS catalogues) in the case of SDSS (2dFLenS).}

\end{table*}

\section{Data}\label{sec:data}

\subsection{KiDS photometric data}\label{sec:kids}

The Kilo-degree Survey (KiDS, \citealt{kids}) is a deep multi-band imaging survey conducted with the OmegaCAM camera (\citealt{omegacam}) which is mounted on the VLT Survey Telescope (\citealt{vst}). This survey uses four broad-band filters ($ugri$) in the optical wavelengths. KiDS has targeted approximately 1350 deg$^2$ of the sky split over two regions, one on the celestial equator and the other in the South Galactic cap. 

KiDS broadband photometry in the optical is supplemented by the VISTA Kilo-degree Infrared Galaxy (VIKING) survey (\citealt{Edge2013}). The VIKING observations of nearly the same regions (by design) with the near infrared filters $ZYJHK_{\rm s}$ significantly increase the wavelength coverage of KiDS, turning the KiDS dataset into a unique wide-field optical+NIR catalogue particularly suitable for cosmological analysis.

In this work we use the fourth KiDS data release (KiDS DR4; \citealt{kuijken2019}) which covers $1006$ deg$^{2}$ of the sky in 1006 tiles superseding the 440 tiles released in KiDS DR3 (\citealt{kids_dr3}) on which \citet{vakili2019} was based. Reduction of the $ugri$ images was performed with the AstroWISE pipeline (\citealt{astrowise}).  
The 1-percentile limiting AB GAaP magnitudes of the survey are  24.8, 25.6, 25.6, 24.0, 24.1, 23.3, 23.4, 22.4, 22.4 in the $ugriZYJHK_{\rm s}$ bands respectively.
The objects present in the final catalogue were detected from the $r$-band images reduced with the THELI pipeline (\citealt{theli2, theli1}).	 For a thorough description of the KiDS data processing, we refer the readers to the data release paper (\citealt{kuijken2019}).



The KiDS data reduction involves a post-processing procedure in which Gaussian Aperture and PSF (GAaP,~\citealt{gaap}) magnitudes are derived (\citealt{kuijken2015}). This procedure is performed in the following way. First, the PSF is homogenized across each individual coadd. Afterwards, a Gaussian-weighted aperture is used to measure the photometry. The size and shape of the aperture is determined by the object's length of the major axis, its length of the minor axis, and its orientation, all measured in the $r$-band. This procedure provides a set of magnitudes for all filters. We refer the readers to \citet{kuijken2015} and \citet{kids_dr3} for a more detailed discussion of the derivation of GAaP magnitudes.
\begin{figure*}
\centering
\includegraphics[width=\textwidth]{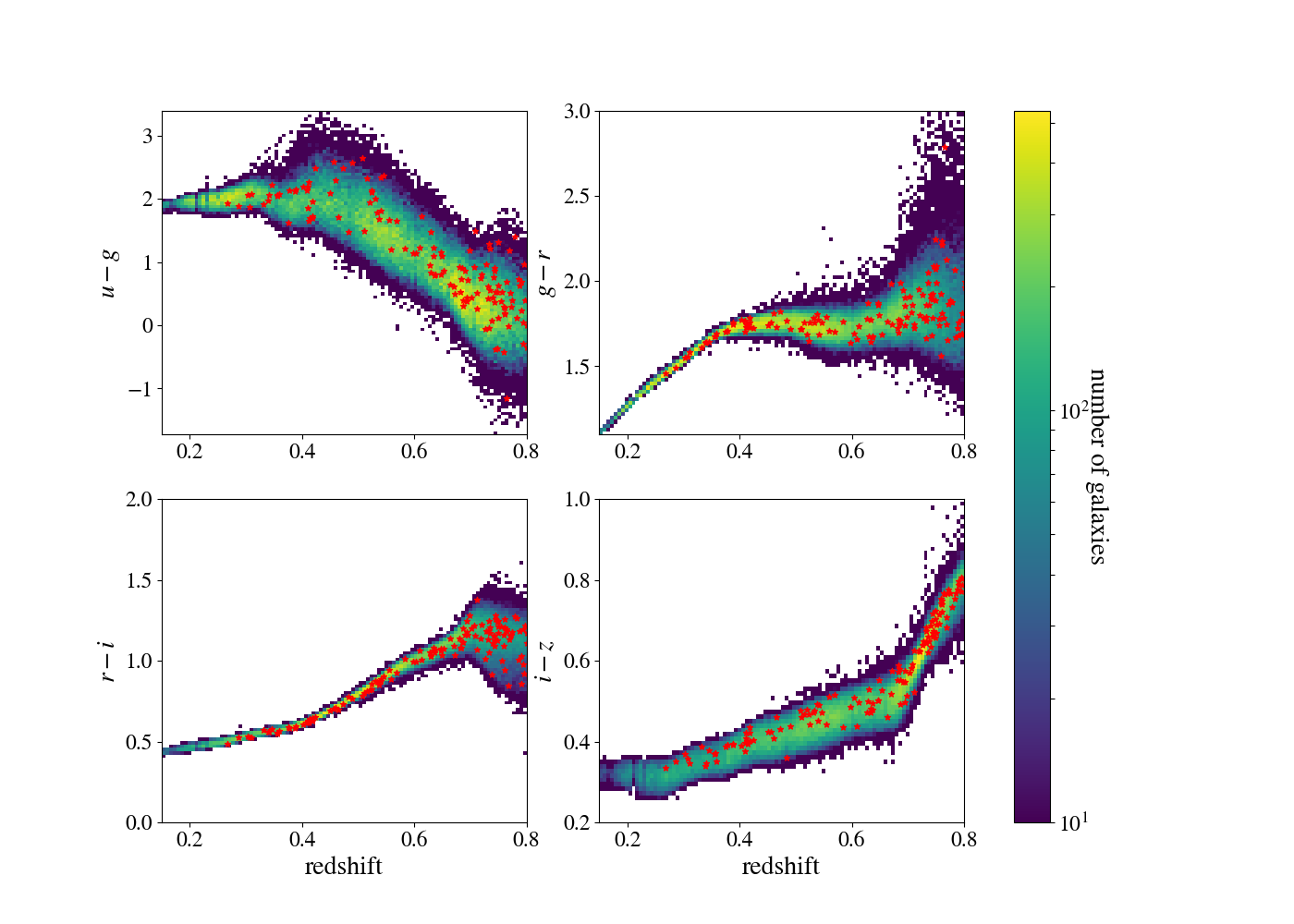}
\caption{Distribution of the selected luminous red-sequence galaxies in colour  space as a function of redshift (colour map). The COSMOS-G10 galaxies are shown as red stars. In each panel, the colour  scale denotes the number density of luminous red galaxies in the colour-redshift space, with yellow corresponding to higher number densities and blue corresponding to lower number densities.} 
\label{fig:cosmos_color}
\end{figure*}
\begin{figure}
\includegraphics[width=\columnwidth]{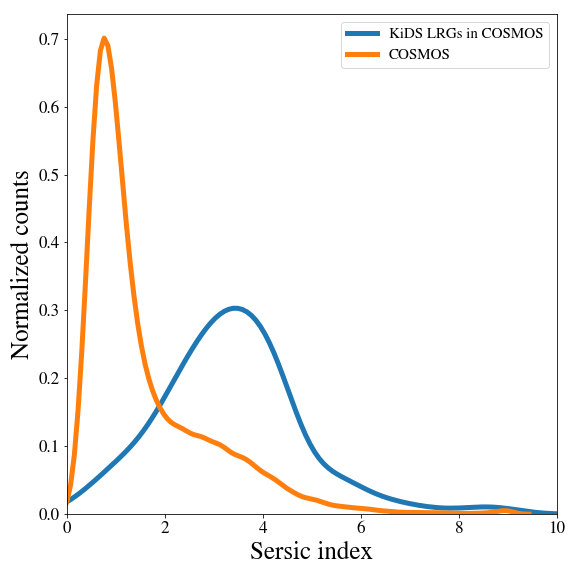}
\caption{Distribution of S\'{e}rsic indices of luminous red-sequence galaxies selected in GAMA-G10 (blue) versus that of all galaxies in the COSMOS Structure and Morphology catalogue (orange). The selected red-sequence galaxies tend to have larger values of S\'{e}rsic indices compared to all galaxies in the COSMOS region.}
\label{fig:cosmos_sersic}
\end{figure}
The magnitudes used in this work are the zeropoint-calibrated and foreground dust extinction-corrected magnitudes. 
The GAaP magnitudes provide accurate colours but underestimate total fluxes of large galaxies. Total fluxes are, however, needed in our LRG selection procedure to derive luminosity ratios.
The magnitude types that provide total fluxes are Source Extractor-based $\mathtt{AUTO}$ magnitudes (\citealt{sextractor}), which are provided in the $r$-band. For the rest of this paper, we work with the $\mathtt{AUTO}$ $r$-band magnitude and GAaP colours. The nine-band photometric catalogue is supplemented by a mask that flags satellite tracks and imaging artefacts such as stellar halos from the images. We use the 9-band catalogues' mask that requires detection of objects in all the 9 bands.



\subsection{Spectroscopic data}\label{sec:spec}

In~\citet{vakili2019} we made use of the spectroscopic data of SDSS DR13 (\citealt{sdss_dr13}), GAMA (\citealt{driver2011}), and 2dFLenS (\citealt{blake2016}). In this work, we also take advantage of the KiDS deep field observation of the COSMOS field. In the COSMOS field we utilize the GAMA-G10 COSMOS spectrocopic data (\citealt{davis2017}), which encompasses a deeper magnitude range, albeit over a much narrower area than the other spectroscopic data considered in this work. The GAMA-G10 catalogue consists of a curation of the redshifts of bright galaxies in the COSMOS region. 
It is important to note that the COSMOS region is not within the KiDS DR4 footprint as it was not observed by VIKING (although it does have KiDS photometry). Instead, KiDS DR3 and VIKING-like\footnote{In the near-infrared photometry of the cosmos region, the CFHTLS $Z$-band is used.} photometric data collected in this area serves as one of the deep photometric redshift calibration samples in KiDS DR4. 


A brief description of these spectroscopic catalogues is provided in Table~\ref{tab:zspec}. For objects with duplicate redshifts in our spectroscopic compilation, we exclude the objects in SDSS that are present in the GAMA catalogue, and we exclude the objects in 2dFLenS that are present in the SDSS or GAMA catalogues, and homogenize the reference frame in which the redshifts are measured\footnote{The redshifts of SDSS DR13 and GAMA galaxies are reported in the heliocentric frame while the redshifts of the 2dFLenS galaxies are reported in the CMB rest frame.}. We note that for the luminous red galaxies, the redshifts obtained by GAMA, SDSS, and 2dFLenS agree on average to $|\delta z|<5\times 10^{-4}$ level with a scatter that increases with redshift. We note that these differences can only mildly impact the uncertainty over the mean values of the redshift distributions.

\begin{figure*}
\begin{tabular}{cc}
\includegraphics[width=0.5\textwidth]{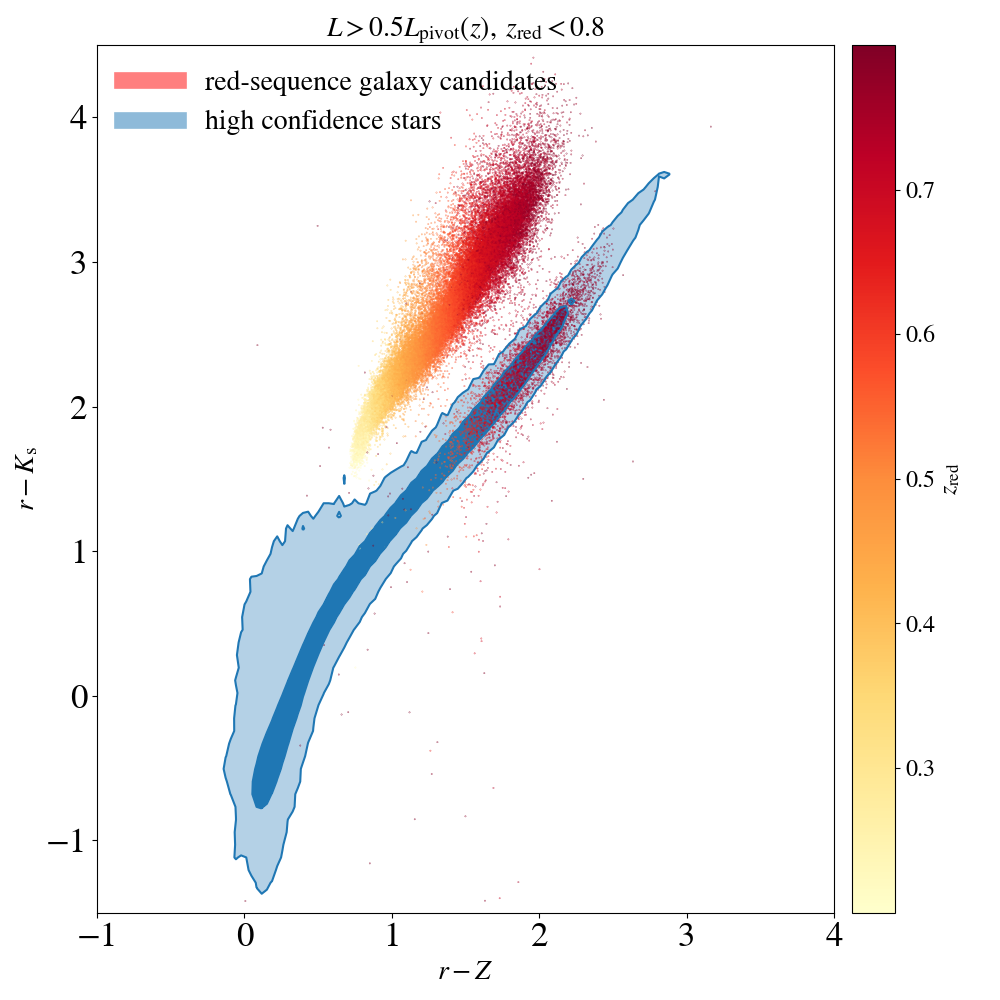}
\includegraphics[width=0.5\textwidth]{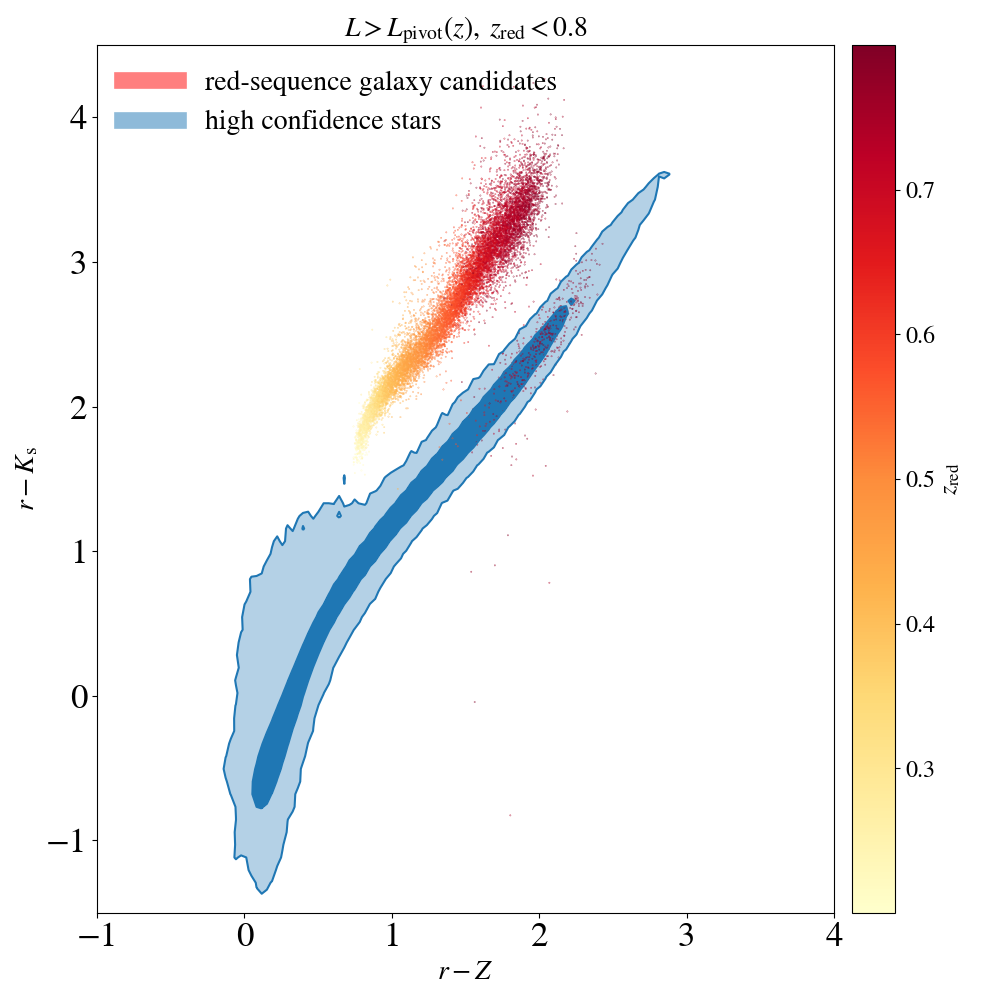}
\end{tabular}
\caption{Demonstration of the use of optical+NIR colours for the identification of likely stellar objects amongst the red-sequence galaxy candidates. 
In each panel, the points colour-coded with redshift show the red-sequence candidates in the $(r-K_{\rm s}) \times (r-Z)$ space, while the blue contours show the 68\% and 95\% confidence regions of the distribution of high confidence stars. Left Panel: At high redshifts ($z_{\rm red}>0.6$), the considerable overlap between the distribution of red-sequence candidates in the dense sample ($L>0.5L_{\rm pivot}(z)$) and that of the high confidence stars becomes clear. Right Panel: In the case of the red-sequence candidates in the luminous sample ($L>L_{\rm pivot}(z)$), the overlap between the two distributions is less apparent.} 
\label{fig:star_galaxy_I}
\end{figure*}

\section{Sample selection}\label{sec:selection}

\subsection{Red-sequence model}

Several aspects of the selection procedure in this work are similar to what has been outlined  in~\citet{vakili2019}. In what follows we describe the main distinctions. Previously, we only utilized the KiDS optical photometry for our red-sequence model. In this work, we also include the VIKING $Z$ band in the red-sequence template.
The added advantage of the $Z$ band is the additional constraining power on the redshifts of the red-sequence galaxies at higher redshifts ($z>0.7$). In principle, one could also include the $YJHK_{\rm s}$ bandpasses of VIKING in the red-sequence model. However, we decided to exclude those bands in the modelling as they would increase the computational cost of selecting the set of seed galaxies (with spectroscopic redshifts) for estimating the parameters of the red-sequence template, and eventually computing the conditional probability of colours conditioned on the redshift and magnitudes for all the objects in the survey.


Our data-driven model of the colours of the red-sequence galaxies (see~\citealt{vakili2019}) is fully characterized with the probability of the colours of red galaxies conditioned on their apparent magnitudes and redshifts: $p(\boldsymbol{c}|m_{\rm r},z)$. This conditional probability is a modelled by a mulitvariate Gaussian distribution. The mean and the covariance of this distribution are parametric functions, via cubic spline, of the apparent magnitudes as well as the redshifts of galaxies. First, we use a mixture of Gaussians to the distribution of spectroscopic galaxies in the colour-magnitude space in thin slices of redshift. Then we estimate the parameters of the parametric functions using the selected seed galaxies.

After evaluating this probability distribution for every object in the survey, we proceed in a similar fashion as described in detail in \citet{rozo2016, vakili2019}. We estimate the best-estimate redshift $z_{\rm red}$ and its uncertainty $\sigma_z$ from the conditional probability distribution $p(z|m_{\rm r}, \boldsymbol{c})$ (see Eqs. 5-7 of \citep{vakili2019}). Then,  we construct two luminosity-threshold samples with the luminosity ratio defined as
\begin{equation}
    \frac{L}{L_{\rm pivot}\left(z\right)} = 10^{-0.4\left(m_{\rm r} - m_{\rm r}^{\rm pivot}\left(z\right)\right)}, \label{eq:lratio}
\end{equation}
in which the characteristic magnitude $m_{\rm r}^{\rm pivot}(z)$ is evaluated using the EZGAL (\citealt{ezgal_paper}) implementation of the \citet{bc03} stellar population model. In the calculation of $m_{\rm r}^{\rm pivot}(z)$ we assume a solar metalicity, a Salpeter initial mass function (\citealt{chabrier2003}), and a single star formation burst at $z_{\rm f} = 3$. Furthermore, this stellar population model is adjusted such that $m_{\rm i}$ = 17.85 at $z=0.2$, matching the magnitude of the redMaPPer cluster galaxies (\citealt{redmap_des, rozo2016}).

The samples are defined by setting a lower bound on the luminosity ratios given by equation~\ref{eq:lratio} and by having a constant comoving density. We define two samples: the dense sample with $L/L_{\rm pivot}(z) > 0.5$ and comoving density of $10^{-3}$ $\mathrm{Mpc}^{-3}h^3$, and the luminous sample with $L/L_{\rm pivot}(z) > 1$ and comoving density of $2.5\times 10^{-4}$ $\mathrm{Mpc}^{-3}h^3$. The comoving densities are calculated assuming a $\Lambda$CDM model with $Planck$ (\citealt{planck2016}) best-fit parameters. We find that the choice of cosmology has no impact on the estimated red-sequence redshifts. 

Using the $\mathtt{LePhare}$ code (\citealt{arnouts1999,ilbert2006}) we derive the stellar masses and the absolute magnitudes of the galaxies in these two samples. For the dense sample, the medians along with the confidence intervals based on the 16th and 84th percentiles of the derived quantities $\log\left[M_{\star}/h_{70}^{-2}M_{\odot}\right]$ and $M_{\rm r}$ are $10.86_{-0.26}^{+0.29}$ and $-22.3_{-0.7}^{+0.6}$ respectively.  
For the luminous sample, the medians along with the 68 percentiles of the same quantities are $11.03_{-0.19}^{+0.23}$ and $-22.7_{-0.5}^{+0.3}$ respectively.

\begin{figure*}
\begin{tabular}{cc}
\includegraphics[width=0.5\textwidth]{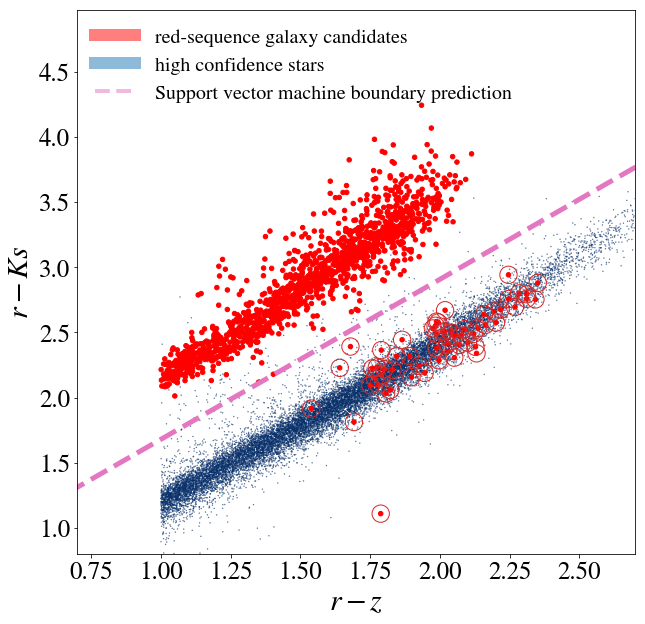}
\includegraphics[width=0.5\textwidth]{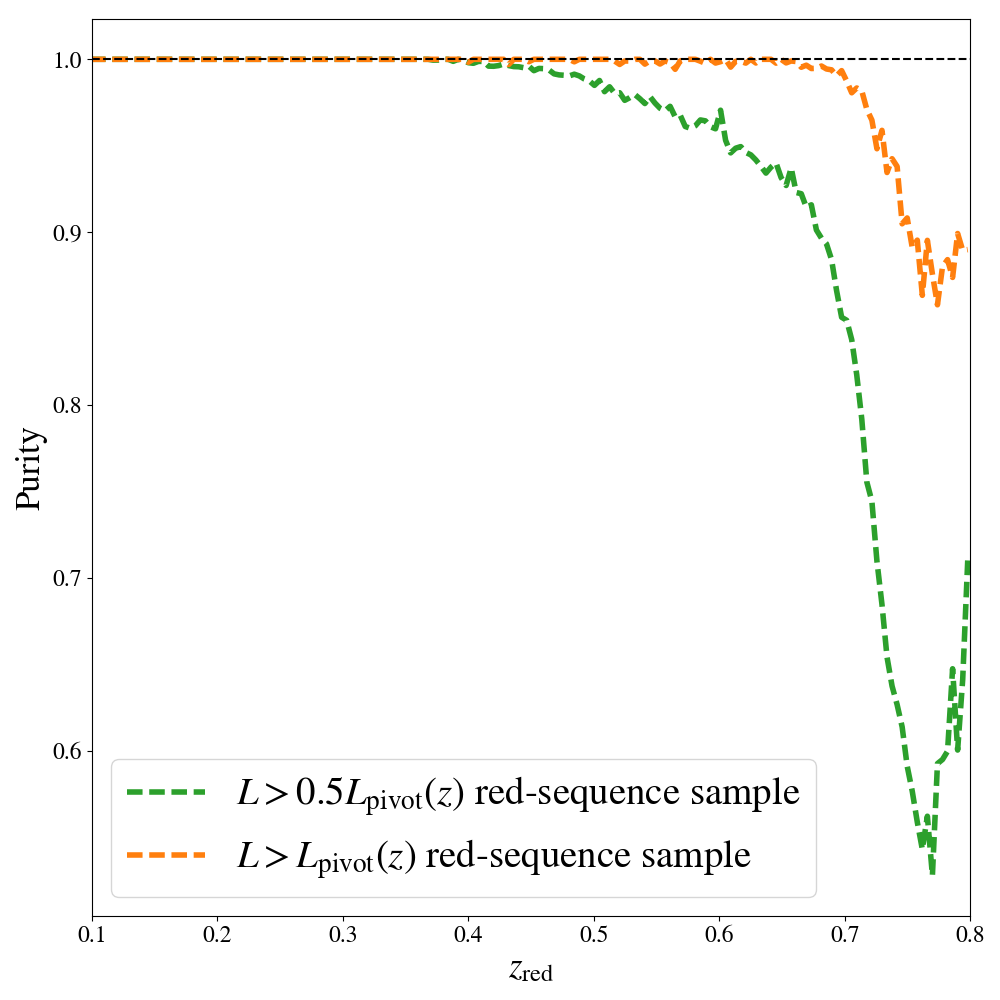}
\end{tabular}
\caption{ Left Panel: At redshifts above $z_{\rm red}>0.4$ red-sequence galaxies (shown in red) and high confidence stars (shown in blue) reside in separated regions of the two-dimensional $(r-K_{\rm s}) \times (r-Z)$ colours. Shown by pink dashed line is the predicted decision boundary between the two classes. The red-sequence candidates falling below the predicted boundary are marked by open circle. These objects are flagged as likely stellar objects in the final catalogue, and thus removed from our large-scale structure analysis. Right Panel: Purity fraction of the dense  (green dashed line) and the luminous (orange dashed line) samples as a function of redshift.}
\label{fig:star_galaxy_II} 
\end{figure*}

For the luminous sample, we show the evolution of the GAaP colours with respect to the estimated red-sequence redshifts in Fig.~\ref{fig:cosmos_color}. The red points show the red-sequence galaxies with $L>L_{\rm pivot}(z)$ in the GAMA-G10 COSMOS field. These galaxies are selected in a consistent manner and hence, they follow the redshift-dependent colour distribution of the luminous red-sequence sample in KiDS DR4. 

Figure~\ref{fig:cosmos_color} offers an intuitive picture of how different colours contribute to determination of the redshifts of red galaxies\footnote{Note that in this intuitive description we have neglected the magnitude dependence of the red-sequence template which plays an additional constraining role in determining the redshifts.}. At low redshifts, the $g-r$ colour rises sharply with increasing redshift. As the 4000 \AA\ break moves between the broadband filters, the $g-r$ colour reaches a relative plateau while the $r-i$ colour starts a rapid increase. At high redshifts however, it is the $i-Z$ colour that shows a higher sensitivity to the redshift of red galaxies. The $u-g$ colour shows a slow and noisy decline considering that red galaxies become fainter in the $u$ filter at higher redshifts. 

We are also interested in the morphological properties of the selected sample, i.e. whether the selected galaxies tend to have morphological parameters associated with elliptical morphologies. We match the red-sequence galaxies in the GAMA-G10 field with the Zurich Structure and Morphology catalogue (\citealt{scarlata2007, sargent2007}). This catalogue contains the best-fit parameters of the Single-S\'{e}rsic GIM2D model applied to the HST ACS imaging data of the COSMOS galaxies. We extract the $\mathtt{SERSIC\_N\_GIM2D}$ column of this catalogue which represents the best-fit S\'{e}rsic index. In Fig.~\ref{fig:cosmos_sersic} the distribution of the S\'{e}rsic indices of red-sequence galaxies in GAMA-G10 is shown in blue, while that of all galaxies in the Zurich catalogue is shown in orange. It is clear that S\'{e}rsic indices of the selected red galaxies in GAMA-G10 tend to have higher values, consistent with the picture that these galaxies are better described by a bulge-dominated morphology common amongst galaxies with old stellar populations.

\subsection{Purity and completeness}\label{sec:purity}

We assess the purity of the sample by inspecting the distribution of the selected objects in the $(r-Z, r-K_{\rm s})$ space. 
In this 2D colour space we focus on the selected red-sequence galaxies and the objects classified as high confidence star candidates in KiDS DR4, i.e. the objects that have\footnote{The $\mathtt{SG\_FLAG}$ parameter makes use of the size-peakiness relation of objects in order to determine whether they can be classified as star or not.} $\mathtt{SG\_FLAG}= 0$.
In Fig.~\ref{fig:star_galaxy_I}, we show the distribution of red-sequence galaxies and high confidence stars in this 2D space. 
The left (right) panel of Fig.~\ref{fig:star_galaxy_I} shows this distribution for the selected objects in the dense (luminous) sample colour-coded by the estimated redshifts. The contours show the 68\% and 95\% of the distribution of high confidence stars in this space. 

As evident in the left panel of Fig.~\ref{fig:star_galaxy_I}, there is some overlap between the distribution of the colours of galaxies in the dense sample with $z_{\rm red}>0.6$ and the distribution of the colours of high confidence stars. In contrast, there is a clear distinction between the colour distribution of objects in the luminous sample and that of the high confidence stars. In both cases, there is a clear gap between the objects labeled as stars and a large majority of the selected red-sequence objects. In the redshift range of  $z_{\rm red}>0.6$ $\sim 40\%$ ($\sim 5\%$) of the objects in the dense (luminous) sample are ambiguous. 

The difference between the purity of the dense and the luminous samples at high redshift arises from the different size-magnitude distributions of the two samples. The objects with higher estimated redshifts tend to be fainter and smaller, making them difficult to distinguish from high confidence stars. The objects in the dense sample with $z_{\rm red}>0.6$ tend to have higher apparent magnitudes and smaller sizes compared to the objects in the luminous sample. This is in line with the findings of~\citet{rozo2016}, according to which the stellar contamination is higher amongst fainter red-sequence objects.

We use Support Vector Machines (hereafter SVM, see~\citealt{cortes1995, cristianini2000, scholkopf2000}) to estimate a decision boundary (a line) that maximizes the margin between the objects in the two classes in 2D space. SVMs are a class of maximum margin classifiers in which a decision boundary is chosen such that the margins between multiple classes are maximized.

It is important to note that we have made an explicit choice of feature engineering for this task. That is motivated by our observation of the gap between the distribution of the two labels in the $(r-Z, r-K_{\rm s})$ plane. Our motivation for using SVM is that in this 2D space, there is a clear margin between the two labels and therefore our choice of a maximum margin classifier for this task is appropriate.


The left panel of Fig.~\ref{fig:star_galaxy_II} shows the predicted decision boundary (the pink dashed line) separating the two red-sequence objects and the high confidence stars. The selected red-sequence galaxy candidates on the right hand side of the decision boundary---shown by red open circles---are likely stellar objects that cannot be differentiated from galaxies with morphological information only. Such objects are removed from the red-sequence samples in order to maximize the purity of the sample. The red-sequence sample purity (impurity) can be quantified as the fraction of red-sequence candidates that lie above (below) the decision boundary shown in the left panel of Fig.~\ref{fig:star_galaxy_II}. The right panel shows the redshift-dependence of the estimated purity of red-sequence objects in the dense (luminous) sample shown in green (orange).

Evidently, the estimated purity of galaxies in the dense sample drops significantly for $z_{\rm red}>0.6$. On the other hand, the purity of the luminous sample remains nearly above 90\% across the entire redshift range $0.1<z_{\rm red}<0.8$. Excluding the contaminants from the dense sample undermines the constant comoving density of this sample for redshifts higher than 0.6. 

\begin{table*}
	\caption{Redshift bin information}
    \centering
	\label{tab:pz}
	\begin{tabularx}{0.82\textwidth}{lcccccccr} 
		\hline
		Redshift bin & Sample & \# objects & $\langle z_{\rm red} \rangle$ & scatter & $m^{\rm phot}_{r, 84\%}$ & $m^{\rm phot}_{r, 99.85\%}$ & $m^{\rm spec}_{r, 84\%}$ & $m^{\rm spec}_{r, 99.85\%}$ \\
		\hline
		$0.15 <z_{\rm red}<0.3$  & $\mathtt{dense}$ & 32225 & 0.241 & 0.014  & 19.86 & 20.52 & 19.69 & 20.2\\
		$0.3  <z_{\rm red}<0.45$ & $\mathtt{dense}$ & 78086 & 0.383 & 0.016 & 20.98 & 21.72 & 20.21 & 21.16 \\
        $0.45 <z_{\rm red}<0.6$  & $\mathtt{dense}$ & 124668 & 0.531& 0.012 & 22.11 & 22.96 & 21.19 & 22.28 \\
        $0.6  <z_{\rm red}<0.8$  & $\mathtt{luminous}$ & 56880 & 0.704 & 0.019 & 22.77 & 23.58 & 21.93 & 23.10 \\
		\hline
	\end{tabularx}
	\tablefoot{The redshift bins (1st column), the mean redshifts (3rd column), their corresponding scatters (4th column, and visualized in Fig.~\ref{fig:zphotscatter}), the 84- and 99.85-percentiles of the GAaP apparent magnitudes in the $r$-band of the selected galaxies in each redshift bin (6th and 7th columns), and the 84- and 99.85-percentiles of the GAaP apparent magnitudes in the $r$-band of the selected galaxies with spectroscopic redshift. The scatter is defined as the scaled median absolute deviation of $\frac{(z_{\rm red} - z_{\rm spec})}{1+z_{\rm spec}}$.}

\end{table*}


\begin{figure}
\includegraphics[width=\columnwidth]{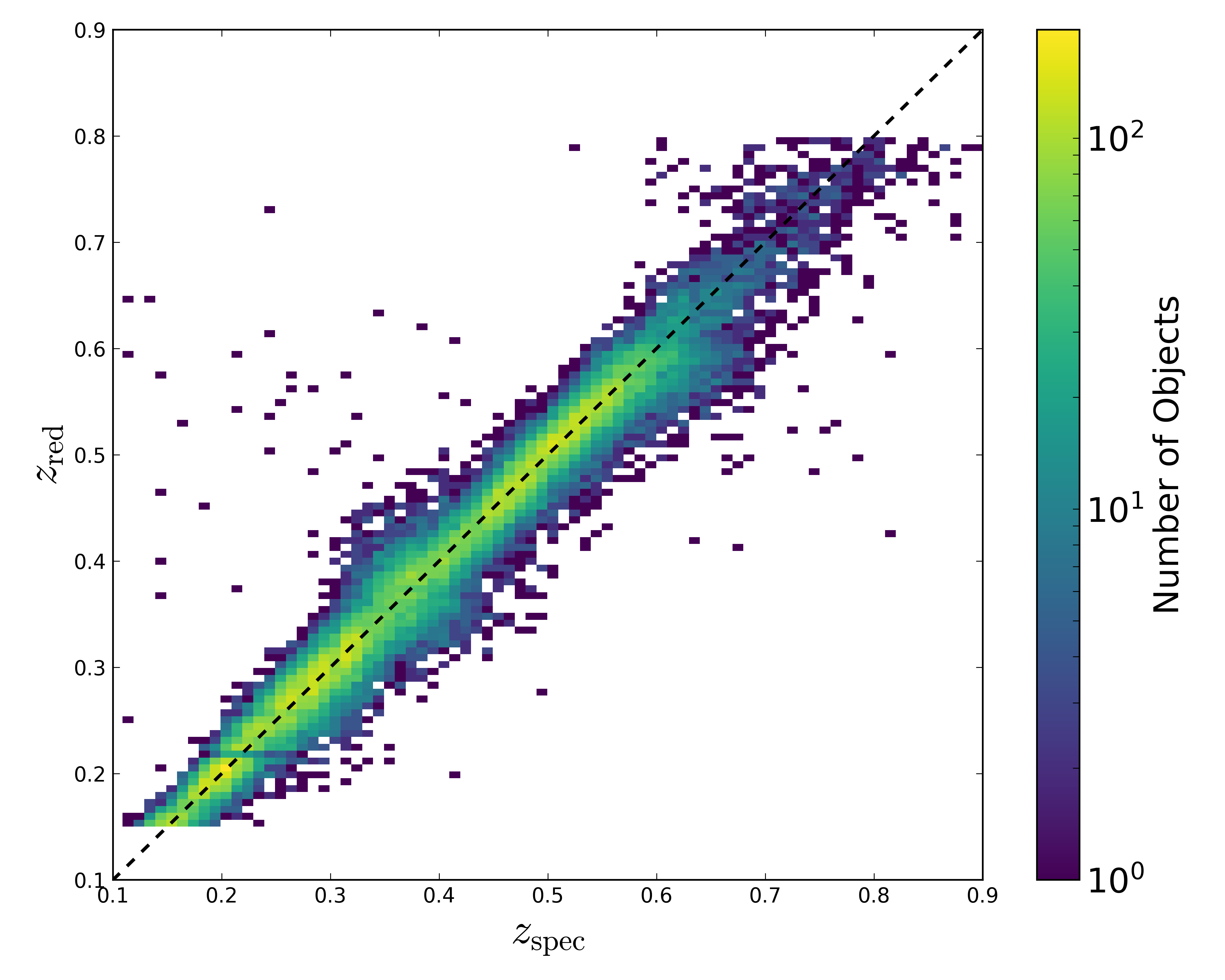}
\caption{Comparison between the estimated red-sequence redshifts and spectroscopic redshifts for galaxies with spectroscopy.} 
\label{fig:zphotscatter}
\end{figure}
\begin{figure}
    \includegraphics[width = \columnwidth]{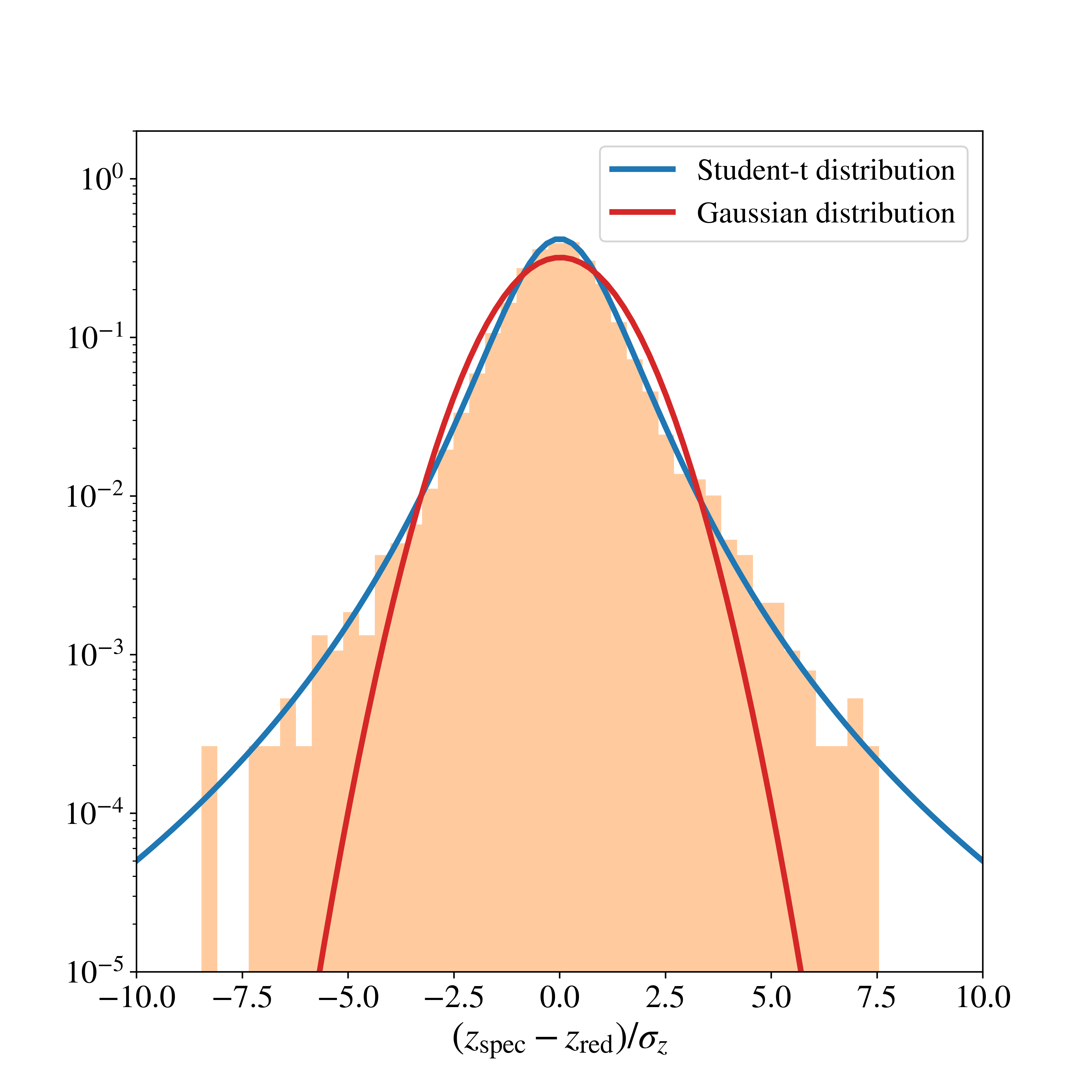}
    \caption{The distribution of the quantity $(z_{\rm spec} - z_{\rm red})/\sigma_z$ is shown in orange, where $z_{\rm red}$ and $\sigma_z$ are per galaxy estimated quantities. Shown in blue (red) is the best-fit Student-t (Gaussian) distribution. A Student t-distribution provides a better description of the long-tails of the redshift distributions of individual galaxies.}
    \label{fig:student-t}
\end{figure}

\begin{table*}
	\centering
	\caption{Photo-z bias divided into four redshift bins and four spec-z data sets.}
	\label{tab:bias}
	\begin{tabularx}{0.73\textwidth}{lcccr} 
		\hline
		Redshift bin & $\rm Bias_{\; GAMA} \times 10^{4}$ &  $\rm Bias_{\; SDSS}\times 10^{4}$ &  $\rm Bias_{\; 2dFLenS}\times 10^{4}$ &  $\rm Bias_{\; G10}\times 10^{4}$ \\
		\hline
		$0.15 <z_{\rm red}<0.3$  & $-5\pm 2 \; (1100)$  & $50\pm 5 \;(8342)$  & $8 \pm 6 \;(847)$  & $40 \pm 4 \;(21)$ \\
		$0.3  <z_{\rm red}<0.45$ & $3 \pm 3 \;(2428)$  & $-10 \pm 5 \;(6449)$  & $0.5 \pm8 \;(1105)$  & $80 \pm 50 \;(112)$ \\
        $0.45 <z_{\rm red}<0.6$  & $50\pm6 \;(7499)$  & $-10 \pm2 \;(1237)$ & $5 \pm9\; (1965)$ & $60 \pm40\;(107)$ \\
        $0.6  <z_{\rm red}<0.8$   & $50\pm40\;(994)$  & $30 \pm10\;(37)$ & $-10\pm30\;(541)$ & $-10 \pm70\;(80)$  \\
		\hline
	\end{tabularx}
    \tablefoot{In each redshift bin, we compare the photo-z biases, summarized with mean and standard deviation, with respect to the four spectroscopic surveys. The number of spectroscopic redshifts available in each redshift bin and spectroscopic data set is shown in parenthesis.}
\end{table*}


Another important factor to take into consideration is the variable depth of the survey in the bands used in the red-sequence model. In the fourth data release, the variable depth is provided by the GAaP limiting magnitudes denoted by $\mathtt{MAG}\_\mathtt{LIM}\_\mathtt{band}$, where $\mathtt{band}=ugriZYJHK_{\rm s}$. We inspect the distribution of the selected objects in a two dimensional space spanned by GAaP magnitude and the GAaP limiting magnitude. In particular, we aim to set the redshift reach of the samples such that the distribution of galaxies in this space is not bounded by the limiting magnitude of the survey. We carry out this investigation for the $ugriZ$ bands. 

For the $griZ$ bands the distribution of galaxies is not limited by the depth of the survey as long as a redshift cut of $z_{\rm red} < 0.6$ and $z_{\rm red} < 0.8$ are applied to the dense and the luminous samples, respectively. 
For the $u$-band however, we note that even after applying the redshift cut of $z_{\rm red} < 0.6$ to the dense sample and $z_{\rm red} < 0.8$ to the luminous sample, both samples are bounded by the depth of the survey. The underlying reason of this limitation is that the red-sequence galaxies become faint in the $u$-band at high redshifts. The possible consequences of this problem are tackled in Sect.~\ref{sec:systematic} where we discuss the various survey properties that can affect the observed number density of galaxies.

\subsection{Photometric redshifts}

For our large-scale structure studies, we construct four redshift bins. 
In order to maximize the signal-to-noise ratio of the clustering as well as of the tangential shear signals we make use of the dense sample as far as the purity and completeness considerations allow us (see Sect.~\ref{sec:purity}). We construct three redshift bins with the dense ($L > 0.5 L_{\rm pivot}(z)$) sample: $0.15<z_{\rm red}<0.3$, $0.3<z_{\rm red}<0.45$, $0.45<z_{\rm red}<0.6$; and finally one redshift bin with the luminous ($L > L_{\rm pivot}(z)$) sample: $0.6<z_{\rm red}<0.8$.

\begin{figure}
    \includegraphics[width = \columnwidth]{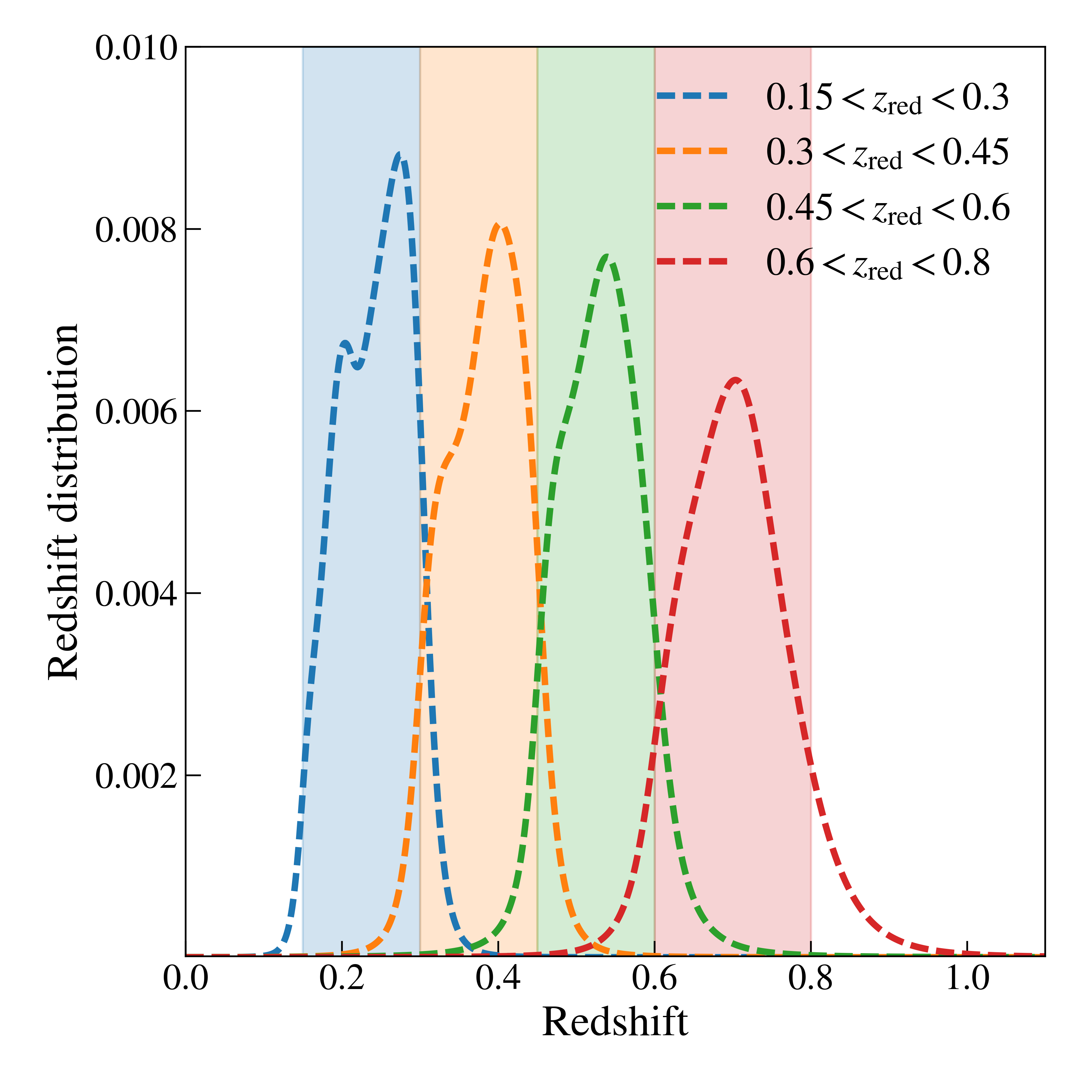}
    \caption{The redshift distributions of the four redshift bins designed for studying the large-scale structure.The shaded regions mark the redshift boundaries used for defining the redshift bins.}
    \label{fig:nzs}
\end{figure}



\begin{table}
	\centering
	\caption{Best-fit Student t-distribution parameters.}
	\label{tab:student-t}
	\begin{tabularx}{0.7\columnwidth}{lccr} 
		\hline
		Redshift bin & $\nu$ & $\mu$ & s \\
		\hline
		$0.15 <z_{\rm red}<0.3$  & 5.89  & 0.055   &  0.907  \\
		$0.3  <z_{\rm red}<0.45$ & 4.36  & 0.005  &  0.898  \\
        $0.45 <z_{\rm red}<0.6$  & 2.92  &  0.02  &  0.820  \\
        $0.6  <z_{\rm red}<0.8$  & 4.03  & -0.015  &  0.875  \\
		\hline
	\end{tabularx}
	\tablefoot{The distribution of the scaled reshift residuals in each bin is modeled by a Student-t probability density. The best-fit parameters of the Student t-distribution are summarized in this table. A lower value of the parameter $\nu$ signals a larger deviation from Gaussianity.}

\end{table}

Table~\ref{tab:pz} summarizes the characteristics of each redshift bin, including the number of objects, the mean redshift $\langle z_{\rm red} \rangle$, the redshift scatter, the 84- and 99.85 percentiles of the $r$-band magnitudes of LRGs and LRGs with spectroscopic redshift. 
The selected objects tend to be fainter than the objects with spectroscopic redshifts. In the last redshift bin, which encompasses the faintest objects, the 84-, and 99.85-percentiles of the r-band GAaP magnitudes in the photometric and spectroscopic samples in the last redshift bin are [22.77, 23.58] and [21.93, 23.1] respectively. This implies that the redshift scatters quoted in Table~\ref{tab:pz} may be optimistic. The robustness of redshifts depend on the accuracy of the red-sequence template, which itself is simply described by a straight line in the colour-magnitude space at each redshift. The template, estimated with brighter galaxies, is generalizable to fainter samples as long as the assumption of the red-sequence template, a ridge-line in the colour-magnitude space, holds. For a sample of galaxies with spectroscopy, comparison between the estimated red-sequence redshifts and the spectroscopic redshifts is shown in Fig.~\ref{fig:zphotscatter}.

The photometric redshift scatter, defined as the scaled median absolute deviation of the quantity $\big(z_{\rm spec} - z_{\rm red}\big)/(1+z_{\rm spec})$, is estimated for each redshift bin. The scatter ranges between 0.012 and 0.019 with the last redshift bin $z_{\rm red} \in [0.6, 0.8]$ having the largest scatter. Furthermore, the slight rise in scatter from the first bin to the second one can be attributed to the transition of the 4000 \AA\ break between the broadband filters in the second redshift bin. Table~\ref{tab:bias} summarizes the photometric redshift biases $z_{\rm red} - z_{\rm spec}$ with respect to the four spectroscopic data sets. The biases are, generally, of order $10^{-3}$ with some scatter between the spectroscopic data sets. We will take this scatter into account in Sect.~\ref{sec:inference} where we estimate the uncertainties on the mean values of the redshift distributions of the four redshift bins.

We define the scaled redshift residuals as the difference between the spectroscopic redshifts and the red-sequence redshifts, divided by the red-sequence redshift uncertainties: $\big(z_{\rm spec} - z_{\rm red}\big)/\sigma_{z}$.
In each redshift bin, we fit the distribution of the scaled residuals with the Normal and the Student-t parametric distributions. The probability density function of a Student-t distribution for a random variable $x$ is given by the following form:
\begin{eqnarray}
f(\tilde{x}) &=& \frac{\Gamma\big(\frac{\nu+1}{2}\big)}{\sqrt{\nu \pi}\Gamma\big(\frac{\nu}{2}\big)} \Big(1 + \frac{\tilde{x}^{2}}{\nu} \Big)^{-\frac{\nu+1}{2}} \\
\tilde{x} &=& \frac{x - \mu}{s},
\end{eqnarray}
where $\Gamma$ denotes the Gamma function, the shape of the distribution is controlled by the parameter $\nu$, the parameter $\mu$ sets the mean of the distribution, and the parameter $s$ scales the width of the distribution. For a sufficiently large value of $\nu$, the Student t-distribution converges to a standard Normal distribution with a mean $\mu$ and a standard deviation $s$. In general, a smaller value of $\nu$ corresponds to a distribution with wider tails.  

Figure~\ref{fig:student-t} shows the distribution of the scaled redshift residuals in the second redshift bin along with the best fit Normal and Student t-distributions. The Student t-distribution provides a better description of the distribution of the redshift residuals. In particular, the tails of the distribution are better modeled by the Student-t whereas the Normal distribution fails to capture the long tails. This implies that the individual redshift probabilities have a longer tail than what a simple Normal distribution suggests.

In each redshift bin summarized in Table~\ref{tab:pz} the distribution of the scaled redshift residuals is modelled by a Student t-distribution specified by the best-fit parameters summarized in Table~\ref{tab:student-t}. The redshift distributions based on this assumption will have longer tails than in the case where the individual distributions are described by a Gaussian density function.
In each $z_{\rm red}$ bin, the Student t-distribution provides a model for $p(z_{\rm true}|z_{\rm red})$ which itself is described by a Student t-distribution with the following shape $\tilde{\nu}$, mean $\tilde{\mu}$, and scale $\tilde{s}$ parameters:

\begin{eqnarray}
\tilde{\nu} &=& \nu, \\
\tilde{\mu} &=& z_{\rm red} + \sigma_z  \mu, \\
\tilde{s} &=& \sigma_z  s,
\end{eqnarray}
where the parameters $(\nu, \mu, s)$ parameters given by Table~\ref{tab:student-t}.

We estimate the redshift distribution of each redshift bin by convolving $dN/dz_{\rm red}$ with $p(z_{\rm true}|z_{\rm red})$ which is equivalent to summing the individual redshift probability distribution functions given by $p(z_{\rm true}|z_{\rm red})$.
Figure~\ref{fig:nzs} shows the redshift distributions of the four redshift bins designed for our galaxy clustering analysis.

\section{Imaging systematics}\label{sec:systematic}

By necessity, the data quality of large galaxy surveys such as KiDS is not homogeneous. The variable survey conditions can potentially affect the observed galaxy density and consequently can bias the cosmological inferences with these galaxy samples (\citealt{ross2012clustering, leistedt2014, leistedt2016mapping, zhai2017clustering, elvin2017, bautista2018sdss, crocce2019dark, DESI_systematic, rezaie2019,heydenreich2020, icaza2020clustering}). In this section, first we describe the imaging systematics considered in our analysis, and then we discuss our mitigation strategy. 

\subsection{Survey properties}\label{sec:sys_list}

\begin{figure*}
\centering
\includegraphics[width = 0.9\textwidth]{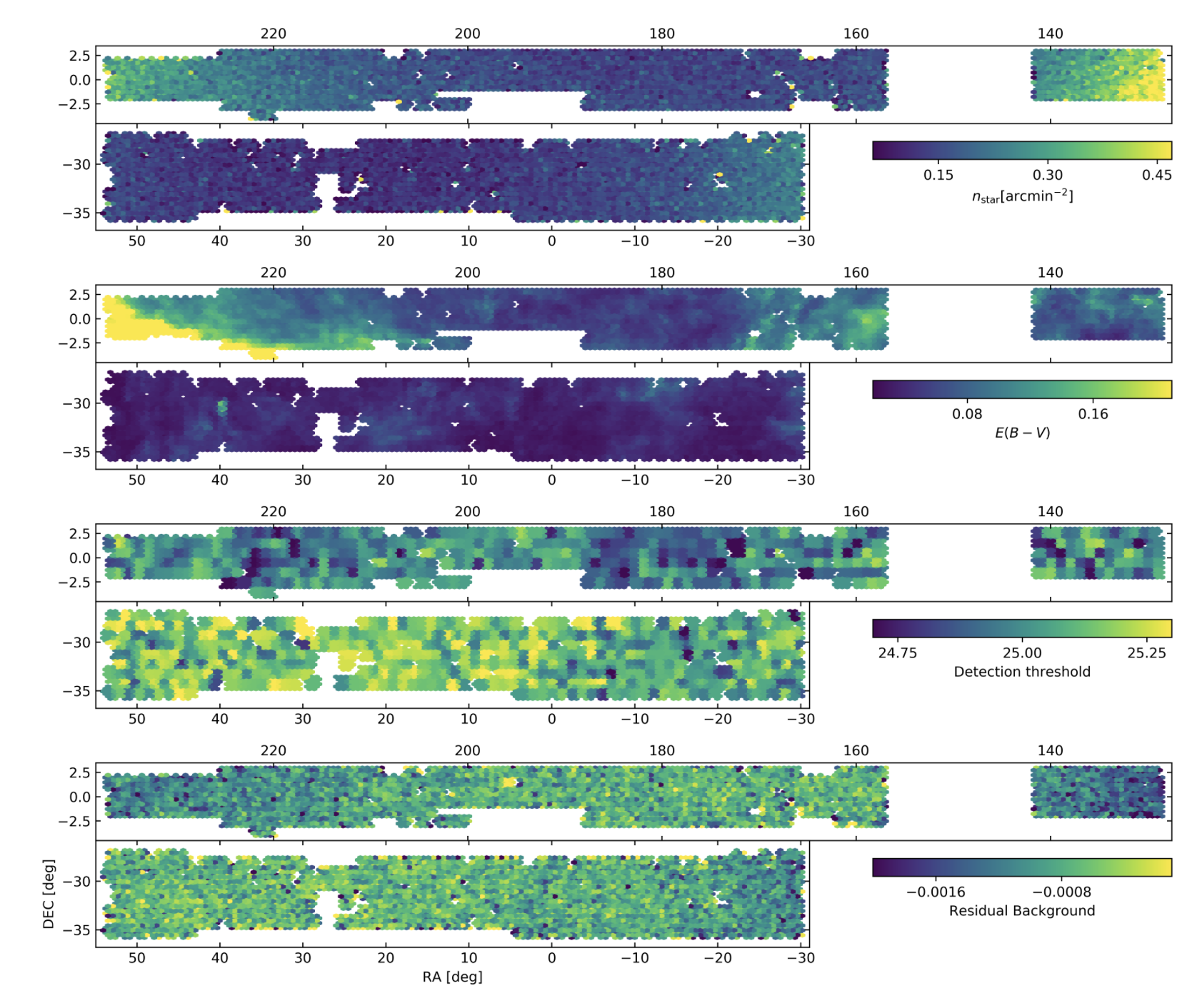}
\caption{ The $\mathtt{HEALPIX}$ maps of the density of GAIA DR2 stars  with $14<G<17$ in the KiDS DR4 footprint (first row), Galactic dust extinction (second row), Detection threshold above background (third row), and Residual Background (fourth row). All maps are generated with $n_{
\rm side} = 256$.} 
\label{fig:scatter_stardens}
\end{figure*}

\begin{figure*}
\centering
\begin{tabular}{ccc}
\includegraphics[width=0.3\textwidth, height =0.3\textwidth]{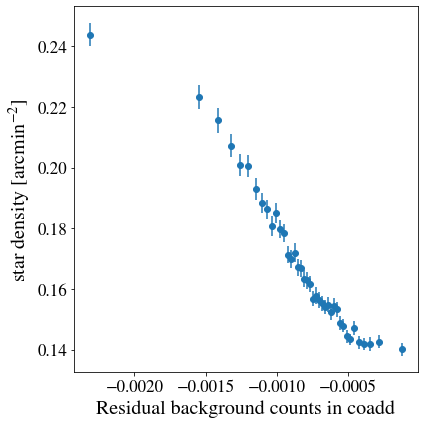}
\includegraphics[width=0.3\textwidth, height =0.3\textwidth]{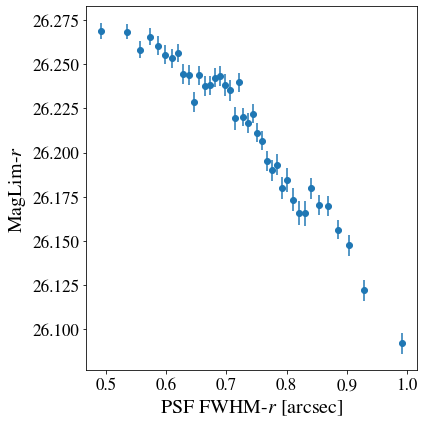}
\includegraphics[width=0.3\textwidth, height =0.3\textwidth]{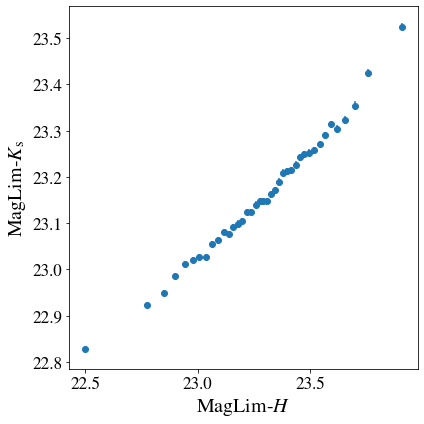}
\end{tabular}
\caption{Demonstration of the correlation between some of the survey properties. 
In each panel, the mean and scatter values of the survey property indicated in the label of the $y$-axis are shown in bins of the survey property indicated in the label of the $x$-axis. The anti-correlation between the residual background counts in the coadds and the stellar number density is evident (\textit{Left}). There is an anti-correlation between the limiting magnitude in the $r$-band and the PSF FWHM $r$-band (\textit{Middle}). Furthermore, there is a correlation between the NIR magnitude limits in the $H$ and the $K_{\rm s}$ bands (\textit{Right})}
\label{fig:sys_sys_correlation}
\end{figure*}

\begin{figure}
\includegraphics[width = \columnwidth, height =\columnwidth]{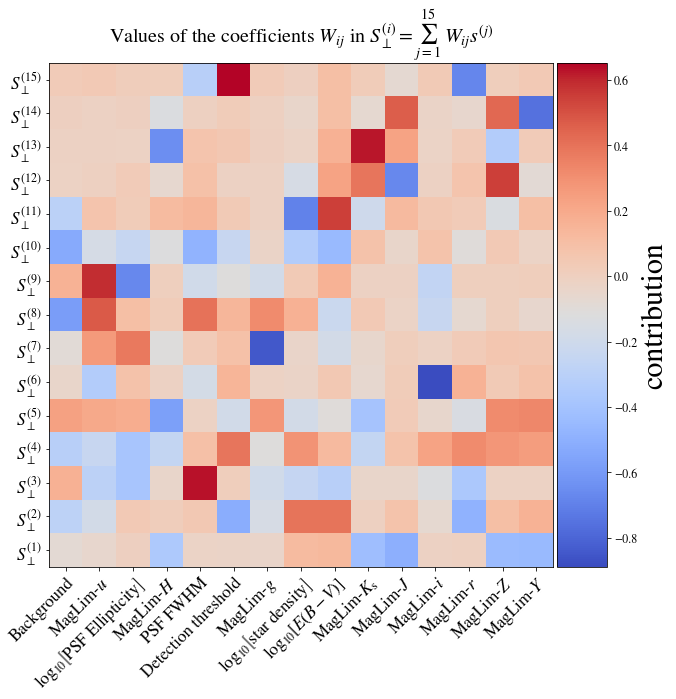}
\caption{Relation between the set of orthogonal survey features $\{S_{\perp}^{(j)}\}_{j=1}^{15}$ (see Eq.~\ref{eq:orth}) and the original survey properties. Some notable features are $S_{\perp}^{(1)}$ dominated by the NIR magnitude limits, $S_{\perp}^{(9)}$ dominated by magnitude limit in the $u$-band and PSF ellipticity, $S_{\perp}^{(11)}$ dominated by the star density and dust extinction.}
\label{fig:pca_components} 
\end{figure}

We consider a range of effects with on-sky variations that can impact the variations of galaxy number densities. In total we take into account 15 survey properties that we further pixelate using $\healpix$ (\citealt{healpix}). We produce the $\healpix$ map of these properties with $N_{\rm side} = 256$ (equivalent to a pixel size of 13.7 arcmin). This choice of map resolution (also adopted by~\citep{ross2012clustering, bautista2018sdss} in clustering measurements of LRGs) is sufficient for mitigating the systematic effects in density variations of a galaxy sample with low number density such as ours. We compute the area of each pixel at higher resolution ($N_{\rm side} = 4096$, equivalent to 0.86 arcmin). This is the resolution at which, the KiDS DR4 mask is provided. That is, the effective areas of large pixels after masking are computed using the unmasked pixels in the high resolution map with $N_{\rm side} = 4096$. 

It is important to note that the detection band in the KiDS photometry pipeline is the $r$-band. Therefore, many of the systematic parameters considered in our analysis are extracted from the $r$-band imaging data. Since we make use of the galaxy GAaP magnitudes and magnitude errors in our red-sequence pipeline, we also include the GAaP limiting magnitudes in our list of imaging systematics.

In what follows, we list the set of survey properties considered in our investigation:

\begin{itemize}

  \item \textbf{Residual background counts in the THELI images}: The background counts at the centroid positions of the objects in the THELI-processed $r$-band detection images. In KiDS DR4, the background count is provided as $\mathtt{BACKGROUND}$. Note that the THELI processed detection images are background subtracted. The $\mathtt{BACKGROUND}$ parameter simply returns the value of the residual sky background at the positions of objects, therefore the background `counts' could be also negative. 

  \item \textbf{Detection threshold above background}: This quantity is measured in units of counts and it is provided in the single-band source list as $\mathtt{THRESHOLD}$. 
  
  \item \textbf{Limiting magnitudes in 9 bands}: The limiting GAaP magnitude attributes are provided in DR4 as $\mathtt{MAG}\_\mathtt{LIM}\_\mathtt{band}$, where $\mathtt{band} = \{u,g,r,i,Z,Y,J,H,K_{\rm s}\}$. 
  For each band the limiting magnitudes are evaluated on an object-by-object basis. At the position of a given object, the limiting GAaP magnitude corresponds to the 1-$\sigma$ GAaP flux error for the aperture of the source. 
  Thus, it depends on the pixel noise---in the Gaussianized image where the GAaP flux is measured---as well as the aperture size. This implies that the limiting magnitudes are indirectly dependent on the full-width-at half-maximum of the point spread function (PSF) in the bandpass as well as the sky background counts. Note that in our red-sequence selection process, we have only used the $ugriZK_{\rm s}$ bands. However since we use the KiDS DR4 9band mask, which requires detection across all 9bands, we also include the limiting magnitudes in the $YJH$ bands in our imaging systematic mitigation.
  
  \item \textbf{PSF full width at half maximum (FWHM) in the $r$-band}: the PSF FWHM in the $r$-band measured in units of arcseconds. The PSF FWHM is calculated using the $\mathtt{PSF\_Strehl\_ratio}$ column in the catalogue.
    
  \item \textbf{PSF ellipticity in the $r$-band}: the KiDS PSF ellipticity in the $r$-band. 
  The PSF ellipticity quantity is computed from the $\mathtt{PSFe1}$, $\mathtt{PSFe2}$ columns in the data. 
  
  \item \textbf{Galactic dust extinction in the $r$-band}: This quantity is provided as $\mathtt{EXTINCTION}\_r$ in the nine-band catalogue of the KiDS DR4 (\citealt{schlegel98, schlafly2011}). 
  
  \item \textbf{Star number density}: We determine the stellar density from the pixelated number density map of bright stars in the second data release of GAIA (DR2, \citealt{gaia0}). This is done by considering the GAIA stars with the $G$-band magnitude between 12 and 17. This is the magnitude range in which the GAIA DR2 $G$-band is complete (\citealt{gaia0,gaia1}). Note that only GAIA DR2 stars that lie in the KiDS footprint are considered in the process of generating the map of stellar number densities. 
  
 
\end{itemize}

The $\mathtt{HEALPIX}$ maps of star number densities, galactic dust extinction, detection threshold, and the residual background counts are shown in Fig.~\ref{fig:scatter_stardens}. The maps of the rest of the survey properties considered in this study are displayed in Figs. 12, 21, and 22 of \citet{kuijken2019}.

\begin{figure*}
\centering
    \includegraphics[width = 0.8\textwidth]{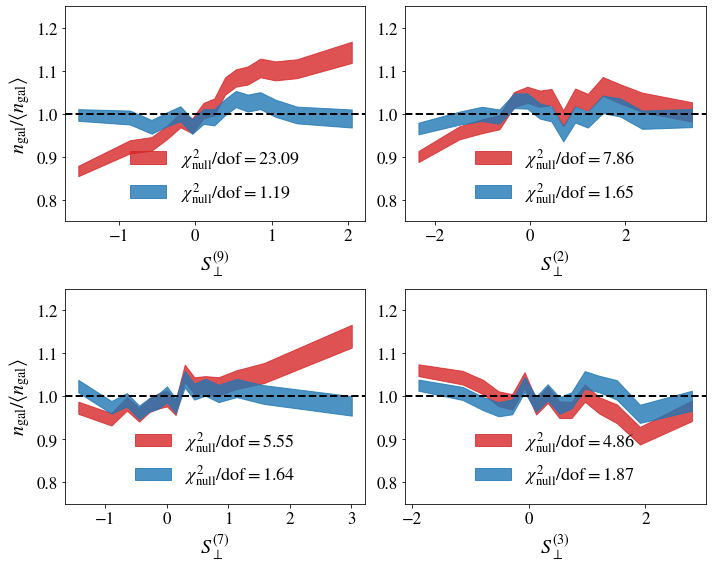}
    \caption{Variation of the galaxy overdensity versus the orthogonal survey property parameters $\{\mathbf{S}_{\perp}\}$ in the third redshift bin $z_{\rm red} \in [0.45, 0.6]$ with (shown in blue) and without (shown in red) the photometric weights included. The deviation of the galaxy over density from unity is quantified by $\chi^{2}_{\rm null}/\mathrm{dof}$ with degrees-of-freedom (dof) = 14. In each panel, the bands show the means and scatters of $n_{\rm gal}/\langle n_{\rm gal}\rangle$ in bins of survey property $S^{(j)}_{\perp}$ shown in the $x$-axis. Note that we have only included the four components of $\{\mathbf{S}_{\perp}\}$ that induce the most significant variations in the observed galaxy number densities. After including the photometric weights, the variation of galaxy densities with respect to survey properties is reduced significantly.}
    \label{fig:sys_ng_correlation}
\end{figure*}


\subsection{The impact of survey properties}


The survey properties considered in our investigation are correlated. Figure~\ref{fig:sys_sys_correlation} demonstrates the mutual information between various survey properties. For instance, there is a strong anti-correlation between the residual background counts in the $r$-band and the stellar number density.
This anti-correlation stems from the tendency of the image processing pipeline to over-estimate, and as a result, to over-subtract the sky background in the fields with higher stellar density. There is also an anti-correlation between the magnitude limit in the $r$-band and the PSF FWHM (middle panel of Fig.~\ref{fig:sys_sys_correlation}). This can be attributed to larger GAaP flux errors for areas of the survey with larger PSF FWHM. 
Conversely, there is a strong correlation between $H$-band and the $K_{\rm s}$ band limiting magnitudes shown in the right panel of Fig.~\ref{fig:sys_sys_correlation}. This stems from the strong relation between the flux errors of these bands. The estimated flux errors are highly correlated in the $ZY$ bands and $HK_{\rm s}$ bands respectively. There is also a strong, albeit with a larger scatter, correlation between the flux errors of all the NIR bands $ZYJHK_{\rm s}$. This mutual information can be attributed to the tiling strategy of the NIR bands. Another possible explanation is the anti-correlation between the limiting magnitudes and the aperture size used for estimating the GAaP flux values. A larger aperture gives rise to a lower limiting magnitude, where the effective photometric aperture is determined by the seeing in each band as well as the degree of detection in the $r$-band.


Given the covariance between some of the survey properties, we inspect the variation of the observed galaxy number densities and the survey properties in the linear basis in which the covariance matrix between the parameters is diagonal. The basis vectors of this space are the eigen-vectors of the covariance matrix of survey properties\footnote{Prior to diagonalization of the covariance matrix of survey properties, we apply a $\log$-transformation to the parameters whose distributions have long tails.}. In such a basis, one can assess the variations between the observed galaxy number density and the different basis vectors independently.

In particular, we inspect the variation of galaxy over-densities $\delta_{\rm gal} = n_{\rm gal}/\langle n_{\rm gal} \rangle$, where $n_{\rm gal}$ is galaxy number density in units of $\rm{arcmin}^{-2}$. As pointed out in Sect.~\ref{sec:sys_list}, the areas pixels in the $\healpix$ maps with $N_{\rm side} = 256$ are calculated using a higher resolution $\healpix$ map of KiDS DR4 mask with resolution of $N_{\rm side}$ = 4096. Any deviation of this quantity from unity as a function of an imaging systematic indicates a non-vanishing impact of that systematic\footnote{Note that alternatively, one can reformulate this by looking at the deviations of $N_{\rm gal}/N_{\rm random}$ (modulo some normalization) from unity, where $N_{\rm random}$ is the number density of a set of uniformly distributed random points across the survey footprint \citep[e.g. ][]{bautista2018sdss, icaza2020clustering}.}.
Let us denote the list of all pixelated survey property parameters by $\{\mathbf{s}_{i} \in \mathbb{R}^{15}\}_{i=1}^{N_{\rm pix}}$ where the subscript $i$ denotes the position of the pixel $i$ on the sky. Furthermore, we transform the set of vectors $\{\mathbf{s}_{i} \in \mathbb{R}^{15}\}_{i=1}^{N_{\rm pix}}$ so that the mean and the variance across each of the 15 dimensions are zero and one, respectively. Afterwards, we transform this 15 dimensional linear basis to a new orthogonal basis in which the covariance matrix of the survey property vectors is diagonal. In this new basis, we represent the list of survey property parameters by $\{\mathbf{S}_{\perp,i} \in \mathbb{R}^{15}\}_{i=1}^{N_{\rm pix}}$ where at each pixel we have:

\begin{equation}
    \mathbf{S}_{\perp} = [S_{\perp}^{(1)}, S_{\perp}^{(2)}, ..., S_{\perp}^{(15)}],
\label{eq:orth}
\end{equation}
where $S_{\perp}^{(j)}$ is the $j$-th systematic vector in the new basis. The relation between the survey properties and the orthogonal features is illustrated in Fig.~\ref{fig:pca_components}. It is clear that the first orthogonal feature is dominated by the near-IR limiting magnitudes. The second feature $S_{\perp}^{(2)}$ is related, with negative signs, to the residual background counts, detection threshold, and the $r$-band limiting magnitude and related, with positive signs, to the star density and the dust extinction.

For the third redshift bin\footnote{For brevity we only show the survey property-density trends in the third bin. The trends in the rest of the redshift bins and the corrections are similar.} ($z_{\rm red} \in [0.45, 0.6]$) the variation of the observed number density versus the survey property parameters $\mathbf{S}_{\perp}$ is shown by the red bands in Fig.~\ref{fig:sys_ng_correlation}, where in each panel, the red bands show the mean and scatter values of the galaxy over-densities in bins of the survey property indicated by the $x$-axis. We have only shown the four most significant variations quantified by the null chi-squared $\chi^2_{\rm null}$, with higher values of $\chi^2_{\rm null}$ corresponding to higher deviations of galaxy densities in relation to the survey properties. For instance, the most significant systematic mode present in density variation of galaxies in the third redshift bin is due to the feature $S_{\perp}^{(9)}$ which itself is dominated by the magnitude limit in the $u$-band. The strong $S_{\perp}^{(9)}$ systematic mode in the third redshift bin (the highest redshift bin constructed from the dense sample) is due to the fact that the $u$-band magnitude distribution of the high redshift galaxies in our sample is limited by the depth of the survey in this band.



It is clear that the galaxy density in our sample correlates with survey properties. Therefore, we mitigate the impact of systematics by introducing a set of photometric weights. This approach has been widely utilized in galaxy clustering analyses: the clustering of LRGs in SDSS BOSS (\citealt{ross2012clustering, ross2017clustering}), galaxy clustering in the Dark Energy Survey (\citealt{elvin2017,crocce2019dark}), DESI legacy survey (\citealt{DESI_systematic}), and finally clustering of LRGs in SDSS-eBOSS (\citealt{bautista2018sdss, icaza2020clustering}). 

Following the work of \citet{bautista2018sdss}, we compute the photometric weights by assuming a relation between the pixelated observed galaxy overdensities and $\delta_{\rm gal} = N_{\rm gal}/\langle N_{\rm gal}\rangle$ and the set of pixelated systematic parameters. In \citet{bautista2018sdss} this relation is assumed to be linear, i.e. $\boldsymbol{\delta}_{\rm gal} = \mathbf{Ws+b} + \mathbf{\mathrm{noise}}$. 
Additionally, we consider two modifications. First, we introduce a set of second-order polynomial features from the original feature space $\{\mathbf{s}\}$. The second-order features consist of all possible combinations of $\{s_{i}s_{j}\}$ as well as single features $\{s_i\}$. These polynomial features are then mapped to the observed galaxy densities via a linear relation. Furthermore, we introduce an $L_{2}$ regularization to this regression problem which is implemented by adding a regularization term $\lambda \sum_{k} W_k^2$ to the least square cost function. The added advantage of this regularization term is that it tends to keep the $\mathbf{W}$ parameters small thereby avoiding overfitting.

In practice, we choose the regularization hyper-parameter $\lambda$ by a $k$-fold cross-validation search in which we explore a wide range of $\lambda$ values from $10^{-4}$ to $10^{5}$. We note that in all the redshift bins, our cross-validation optimization procedure prefers a heavy regularization in which a very large value of $\lambda \in [10^3-10^5]$ is favored. The advantage of this approach over that of \citet{ross2017clustering} is that we do not assume that there is no correlation between the systematic parameters.

The prediction of this model, once applied to the pixelated systematic maps, provides a set of photometric weights that remove the systematically induced variations in the galaxy number density. The photometric weights are obtained by taking the inverse of the prediction of the model. Figure~\ref{fig:sys_ng_correlation} demonstrates how the photometric weights derived from our framework can help reduce the systematic trends seen in the observed galaxy number densities. In Fig.~\ref{fig:sys_ng_correlation}, the density-correlations are displayed after taking into account the photometric weights (shown in blue) and without the photometric weights (shown in red). We note that the reduced $\chi^{2}$ improves significantly once the photometric weights are taken into consideration. 

We also investigated the two-point cross-correlations of the galaxy number density and the orthogonal systematic parameters $\{\mathbf{S}_{\perp, i}\}_{i=1}^{N_{pix}}$ as a function of angular separation in our galaxy sample. We find that the cross correlations, before and after including the photometric weights, are consistent with zero, albeit with slight improvements once the photometric weights are taken into account.


Alternatively, one can use self organizing maps (SOM, \citealt{kohonen1997}) for learning the systematic galaxy density modes due the variable survey properties and then generating a set of `organized randoms' mimicking the galaxy depletion pattern across the survey footprint. We have also tested this method and we found that this method works best in correcting the systematic depletion in a galaxy sample with a higher number density than in our study. This approach is being pursued by Johnston et al (in prep) to mitigate the systematic biases in clustering of galaxies in the KiDS DR4 bright sample (Bilicki et al. in prep.).

We have not explored the effect of various observing conditions on the distribution of derived physical properties of the galaxies. Such effects can in principle generate systematic on-sky variations of the estimated host halo mass of galaxies in our sample which can subsequently complicate the cosmological analysis. This problem is exacerbated in a galaxy survey covering a larger area, due to significant reduction in statistical uncertainty, and can only be taken into account through a careful forward model approach which is outside the scope of this paper.


\section{Galaxy Clustering}\label{sec:clustering}

\subsection{Theory}

Assuming a local deterministic linear galaxy bias, the galaxy overdensity $\delta_g$ is related to the matter overdensity $\delta_m$ through a linear relation: $\delta_g = b_g \delta_m$, where the parameter $b_g$ is the linear bias parameter. Such assumption is expected to hold on sufficiently large scales \citep[e.g.][]{Kravtsov1999, marian2015}, whereas on small scales, the nonlinear structure formation is best described by the more sophisticated halo model \citep[e.g. ][]{berlind2002,cooray2002,zehavi2011,hand2017,vakili_hahn}.

Given a nonlinear matter power-spectrum $P_{\rm NL}(k,z)$, and a galaxy population with linear bias of $b_g$ and redshift distribution $n_g(z)$, one can predict the angular two-point correlation function $w_{g}(\theta)$. Under the assumption of flat universe, and the Limber- and flat-sky approximations (\citealt{limber1961, loverde2008, kilbinger2017, kitching2017}), the angular clustering $w_g(\theta)$ is given by:

\begin{eqnarray}
w_{g}(\theta) &=& b_g^2 \int_{0}^{\infty} \frac{ ldl}{2\pi}J_{0}(l\theta)  \nonumber \\ 
            &\times& \int_{0}^{\chi_{\rm H}} d\chi_{\rm c} \left[\frac{n_{g}(z)\frac{dz}{d\chi_{\rm c}}}{\chi_{\rm c}}\right]^{2}P_{\rm NL}\left(\frac{l+1/2}{\chi_{\rm c}}; z\right),                  
\label{eq:clustering_theory}
\end{eqnarray}
where $\chi_{\rm c}$ is the comoving distance to redshift $z$, $\chi_{\rm H}$ the Hubble distance, and $J_{0}$ the 0-th order Bessel function of the first kind. The nonlinear power spectrum can be calculated using different recipes such as the halo model \citep[e.g. ][]{takahashi2012, mead2015, smith2019}, or emulation \citep[e.g.][]{emu2014}. Hereafter we use the Core Cosmology Library ($\mathtt{CCL}$, \citealt{ccl2019, ccl_code}) to compute the theoretical predictions of angular clustering. We use the~\citet{takahashi2012} model as it is capable of predicting the matter clustering in the linear and quasi-linear regimes (i.e. $\chi_{\rm c}>8 \; \mathrm{Mpc}/h$) considered in this study.

We calculate the theoretical prediction (eq. \ref{eq:clustering_theory}) for all four redshift bins in our galaxy sample with their corresponding linear bias parameters $\{b^{(i)}_{g}\}_{i=1}^{4}$ and redshift distributions $\{n^{(i)}_{g}(z)\}_{i=1}^{4}$.

Assuming a fixed cosmology, we aim to estimate the linear bias parameters $\{b^{(i)}_{g}\}_{i=1}^{4}$ by fitting the theoretical prediction (\ref{eq:clustering_theory}) to the angular clustering of our LRG samples. It is worth noting that one major source of systematic error in this theoretical prediction is the uncertainty on the mean of the redshift distributions $\{n_{g}^{(i)}(z)\}_{i=1}^{4}$. In order to mitigate its impact, we assume that the estimated redshift distribution at each redshift bin $n_{g}^{(i)}(z)$ is effectively given by shifting the underlying unbiased redshift distribution $n_{g, \rm true}^{(i)}(z - \delta z^{(i)})$, where the parameter $\delta z^{(i)}$ is the uncertainty on the mean of the redshift distribution of $i$-th redshift bin. In Sect.~\ref{sec:inference} we will 
discuss how the uncertainty over the mean of the redshift distribution can be estimated. When reporting the constraints on the galaxy bias parameters, we marginalize over the redshift uncertainties. 

\begin{figure*}
\begin{center}
\includegraphics[width = 0.7\textwidth]{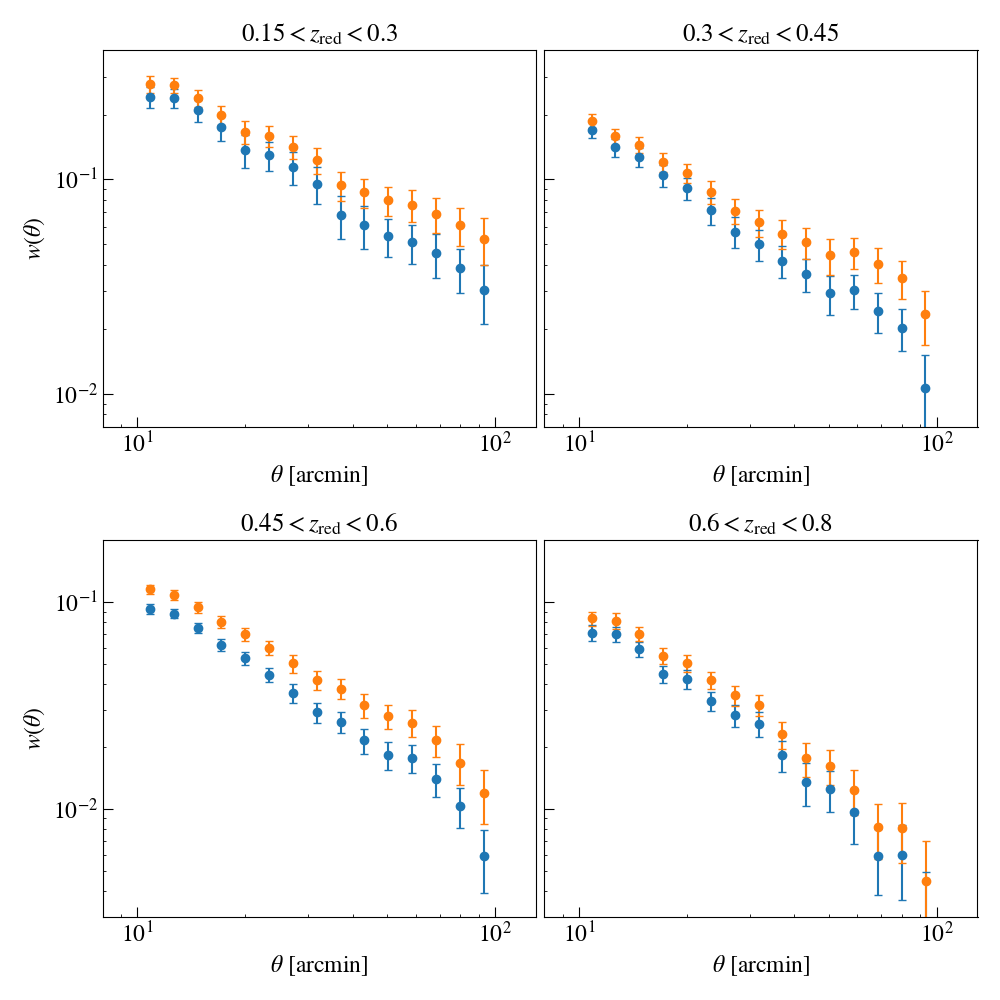}
\caption{Clustering measurements for the four redshift bins. In each panel, the blue (orange) data-points correspond to the correlation functions computed with (without) the photometric weights designed to remove survey-related systematic density variations.}\label{fig:xi} 
\end{center}
\end{figure*}

\subsection{Clustering measurements}\label{sec:measurement}

The galaxy two-point correlation function is the excess probability, compared to random, of finding a pair of galaxies within a given angular or physical separation. Given that we do not know the exact redshifts of the galaxies, we focus on computing the angular correlation function which can be obtained by performing pair-counts in angular bins perpendicular to the line of sight. We measure the angular clustering using the Landy-Szalay estimator (\citealt{landy}):
\begin{equation}
    \widehat{w}(\theta) = \frac{DD-2DR+RR}{RR},
\label{eq:landy}
\end{equation}
where $DD$ denotes the number of galaxy pairs within an angular separation bin centered at $\theta$, $DR$ denotes the number of galaxy-random pairs, and $RR$ denotes the number of random pairs. The random points are uniformly distributed within the survey footprint.

As discussed in Sect.~\ref{sec:systematic}, in each redshift bin a photometric weight is assigned to each galaxy depending on its position on the sky. These weights are derived such that the on-sky variations of galaxy overdensity due to survey properties are mitigated. We compute two sets of angular clustering measurements, one with the photometric weights, and another without. In the presence of weights the $DD$ and $DR$ pair count calculations are modified in the following way:
\begin{eqnarray}
    DD(\theta) &=& \sum_{i=1}^{N_{\rm galaxy}}\sum_{j=1}^{N_{\rm galaxy}}\omega_{i}\omega_{j}\Theta_{ij}(\theta), \\
     DR(\theta) &=& \sum_{i=1}^{N_{\rm galaxy}}\sum_{j=1}^{N_{\rm random}}\omega_{i}\Theta_{ij}(\theta),
\end{eqnarray}
where $\Theta_{ij}(\theta) = 1(0)$ when a galaxy-galaxy, or a galaxy-random pair in the case of $DR$, (indexed by $i,j$) are (are not) within the angular bin centered on $\theta$, and $\omega_{i}$ is the photometric weight associated with the $i$-th galaxy.

The angular clustering measurements of LRGs in four redshift bins are displayed in Fig.~\ref{fig:xi}, with the first three bins encompassing the galaxies in the dense $(L> 0.5L_{\rm pivot}(z))$ sample and the last bin ($0.6<z<0.8$) encompassing the galaxies in the luminous $(L> L_{\rm pivot}(z))$ sample. The clustering signal estimated with (without) the photometric weights is shown in blue (orange). The correlation functions are measured in 15 logarithmically-spaced bins in the range $ 10\leq \theta \leq 100 \; \mathrm{arcmin}$. Although the lowest angular limit of 10 arcmin is lower than the systematic map resolution of 13 arcmin, the scale cuts (see Sec.~\ref{sec:scale_cut}) that will be applied to our clustering data vectors will be larger than both of these angular scales.

We estimate the measurement uncertainties using the jackknife resampling method (\citealt{norberg2009,oliver2016,singh2017,shirasaki2017}). 
In this method, the KiDS survey footprint is first divided into $N_{\mathrm{JK}}=100$ contiguous jackknife subsamples of approximately equal area\footnote{The segmentation of KiDS DR4 footprint into $N_{\mathrm{JK}}$ is done with the $K$-means algorithm. We made use of an implementation of this algorithm designed to handle RA and DEC coordinates on the sky (\hyperlink{kmeans\_radec}{https://github.com/esheldon/kmeans\_radec}).}
For each subsample $k\in\{1,...,N_{\mathrm{JK}}\}$, the clustering data vector $\boldsymbol{w}_{g}^{(k)}$
is measured by dropping the $k$-th subsample and estimating the clustering signal from the rest of the survey area. Note that the vector $\boldsymbol{w}_{g}^{(k)}$ contains the correlation function measured in all the 15 angular bins considered in this study. The jackknife estimator of the covariance matrix is then given by:
\be 
\widehat{C}_{\rm JK} = \frac{\njk - 1}{\njk}\sum_{k=1}^{\njk}\big(\dk-\dbar\big)^{T}\big(\dk-\dbar\big), 
\label{eq:jk}
\ee
where $\dbar$ is the mean of all $\boldsymbol{w}_{g}^{(k)}$ vectors.

Since the covariance matrix is estimated from the jackknife method with a finite number of jackknife subsamples, our estimate of the covariance matrix and its inverse are noisy. The unbiased estimate of the inverse covariance matrix is related to the inverse of the estimated jackknife covariance matrix $\hat{C}_{\rm JK}$ with the Anderson-Hartlap-Kaufman (\citealt{Kaufman1967, hartlap2007}) debiasing factor:

\be
\widehat{C^{-1}} = \frac{\njk - N_{\rm d} - 2}{\njk - 1} \widehat{C}^{-1}_{\rm JK},
\label{eq:hartlap}
\ee
where $\njk$ is the number of jackknife subsamples and $N_{\rm d}$ is the number of bins, in the clustering measurements, that enter the likelihood function evaluation. Since we remove the nonlinear scales from our likelihood analysis, only the data points that pass the cuts determine the number of data points $N_{\rm d}$ appearing in Equation~\ref{eq:hartlap}.


\begin{figure*}
\centering
\includegraphics[width=0.9\textwidth]{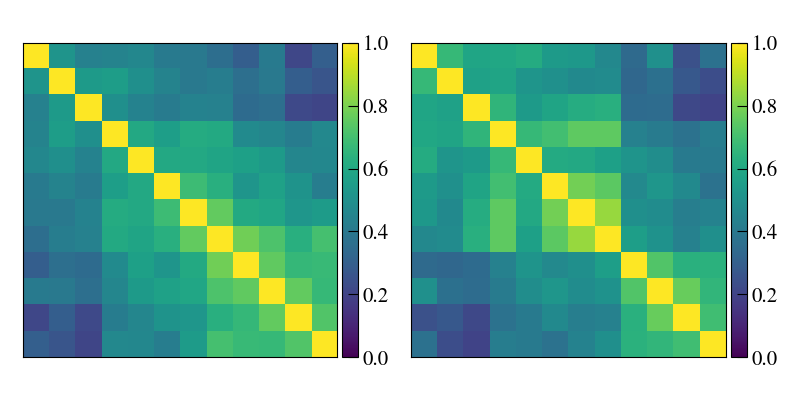}
\caption{Illustration of our blinding scheme based on modifying the covariance matrix. Two correlation matrices corresponding to the angular clustering measurements in the last redshift bin: the matrix shown in the left panel is the original correlation matrix derived from the jackknife covariance matrix computed in Sect.~\ref{sec:measurement}, while the matrix shown in the right panel is constructed with the method of~\citet{sellentin2019} such that the posterior probability over the galaxy bias parameter is shifted.}
\label{fig:blind}
\end{figure*}

\subsection{Inference setup}\label{sec:inference}

\subsubsection{Parameters}

In order to fit the theoretical model of angular clustering to the data, we need to clarify our choices of parameters. Assuming a fixed cosmology, we estimate the linear galaxy bias (using equation~\ref{eq:clustering_theory}) of the red galaxies in the redshift redshift bins. We estimate the bias parameters by marginalizing over the photometric redshift uncertainty parameters.  
Furthermore, we choose a flat LCDM model assuming $\Omega_m = 0.25, \;\Omega_{\Lambda} = 0.75, \; \Omega_b = 0.044, \; \sigma_{8} = 0.8, \; n_s = 0.95, \; h = 0.7$, which is the input cosmology of the MICE suite of cosmological simulations (\citealt{MICE1}). We picked this set of cosmological parameters for internal comparison purposes (Fortuna et al in prep). However, given that the amplitude of galaxy clustering depends on the amplitude of the power spectrum, growth factor, and galaxy bias, we expect our bias constraints in this work to depend on the assumed cosmology.






We adopt the following priors on the model parameters. For the galaxy bias parameters, we assume a uniform prior with a lower bound of 1 and an upper bound of 3. For the prior distribution of the photo-z shift parameters $\{\delta z _{i}\}_{i=1}^{4}$ we assume a Gaussian distribution with zero mean and a dispersion which we estimate in the following way. We assume that there are two major contributions to the uncertainty of the mean of the redshift distributions. The first contribution is the spatial sample variance which we compute using the jackknife resampling method. The second contribution is estimated by computing the covariance between the photo-z biases with respect to the four spectroscopic redshift surveys considered in this study (see Table~\ref{tab:bias}). Combining these two sources of uncertainty will provide us with an estimate of the prior distribution over the photo-z shift parameters:
\begin{equation}
    \delta z_{i} \sim \mathcal{N}\big(0, \sigma_{\delta z_{i}}\big),
\end{equation}
where $\sigma_{\delta z_{i}}$ is $2.4\times 10^{-3}$, $3.2\times 10^{-3}$, $2.3\times 10^{-3}$, and $4.6\times 10^{-3}$ for the first, second, third, and fourth redshift bins respectively. 

\subsubsection{Blinding}

In order to avoid confirmation bias we adopt the blinding scheme introduced in \cite{sellentin2019}. In this approach the blinded element of the inference pipeline is the inverse covariance matrix as opposed to the catalogues (\citealt{hendrick2017}), photo-z distributions (\citealt{hendrik2020}), or the correlation functions (\citealt{muir2019}). 

The estimated inverse covariance matrix is changed by multiplication of a diagonal matrix to the Cholesky decomposition of the estimated inverse covariance matrix\footnote{The Cholesky decomposition of a symmetric positive-definite matrix $A$ is a matrix factorization that can be expressed as $A = LL^{\rm T}$, where $L$ is a unique lower triangular matrix.}. The diagonal matrix is chosen such that the posterior distributions derived from the new inverse covariance matrix are shifted with respect to those estimated from the fiducial inverse covariance matrix. Transformation of the inverse covariance matrix is given by the following set of equations:

\begin{eqnarray}
\widehat{C^{-1}} &=& L^{T}L \\ \label{eq:unblinded}
             &\rightarrow& L^{T}B^{T}BL, \label{eq:blinded}
\end{eqnarray}
where $L^{T}L$ is the Cholesky decomposition of the original inverse covariance matrix, $B$ is the diagonal matrix responsible for shifting the peak of the posterior probability distribution over the galaxy bias parameter, and $L^{T}B^{T}BL$ is the blinded inverse covariance matrix. The elements of the diagonal matrix $B$ are given by
\begin{equation}
B_{ii} = \frac{e_i}{\tilde{e}_i},\label{eq:bii}
\end{equation}
with the vectors $\mathbf{e}$ and $\mathbf{\tilde{e}}$ defined in the following way:
\begin{eqnarray}
e_i &=&  L^{T} \left[\widehat{\mathbf{w}} - \mathbf{w}_{\rm}(b_{\rm fid}) \right] \label{eq:ei} \\
\tilde{e}_i &=& L^{T} \left[\widehat{\mathbf{w}} - \mathbf{w}_{\rm}(\tilde{b}) \right] \label{eq:eni}
\end{eqnarray}

In Eqs~\ref{eq:ei} and~\ref{eq:eni}, $\widehat{\mathbf{w}}$ denotes the clustering measurement data vector, while $\mathbf{w}_{\rm}(\tilde{b})$ and $\mathbf{w}_{\rm}(b_{\rm fid})$ denote the clustering model data vectors evaluated at a fiducial galaxy bias parameter $b_{\rm fid}$ and at a perturbed bias parameter $\tilde{b}$ respectively. The equations~\ref{eq:ei} and~\ref{eq:eni} are designed such that the model $\mathbf{w}(\tilde{b})$ under the blinded inverse covariance matrix given by Eq.~\ref{eq:blinded} achieves the same $\chi^2$ goodness-of-fit as the model $\mathbf{w}(b_{\rm fid})$ under the original inverse covariance matrix given by Eq.~\ref{eq:unblinded}. 
For our dense (luminous) sample we set the parameters $b_{\rm fid}$ and $\tilde{b}$ to 1.8 (2.1) and 1.78 (2.08) respectively.

Figure~\ref{fig:blind} demonstrates the blinding scheme, with the two panels showing the correlation matrix of the clustering measurement corresponding to the last redshift bin before and after blinding. After finalizing all the steps of our analysis, we repeat the last step of our investigation with the original covariance matrices. The difference between our galaxy bias constraints before and after blinding is described in Appendix~\ref{A:blind}. We have not incorporated any invariance of the width of the target posterior pdfs, which can lead to seemingly undesirable feature such as the first panel of Fig.~\ref{fig:blind_vs_unblind_post}. In practice however, this implementation serves our purpose of avoiding confirmation bias.

\subsubsection{Scale cuts}\label{sec:scale_cut}

The theoretical model summarized in equation~\ref{eq:clustering_theory} fails to capture the full complexity of the galaxy-matter connection on small scales as it relies only on a simple linear deterministic treatment of galaxy bias. Therefore, we decided to apply a conservative cut on the comoving scales considered in our theoretical modeling of the clustering signal. In particular, we adopt a cut at a comoving scale of 8 $\mathrm{Mpc}h^{-1}$ which translates to a minimum angular scale (hereafter denoted by $\theta_{\rm min}$) of 39.8 arcmin for the first bin, 25.9 arcmin for the second bin, 19.3 arcmin for the third bin, and finally 15.2 arcmin for the last redshift bin. The comoving distance is converted to angular scales assuming the flat $\Lambda$CDM cosmology discussed above. Furthermore, the parameter $\theta_{\rm min}$ in each redshift bin is calculated from dividing the minimum comoving scale of 8 $\mathrm{Mpc}h^{-1}$ by the comoving distance at the mean redshift of the redshift bin under consideration.

\begin{figure*}
\centering
\includegraphics[width=0.9\textwidth]{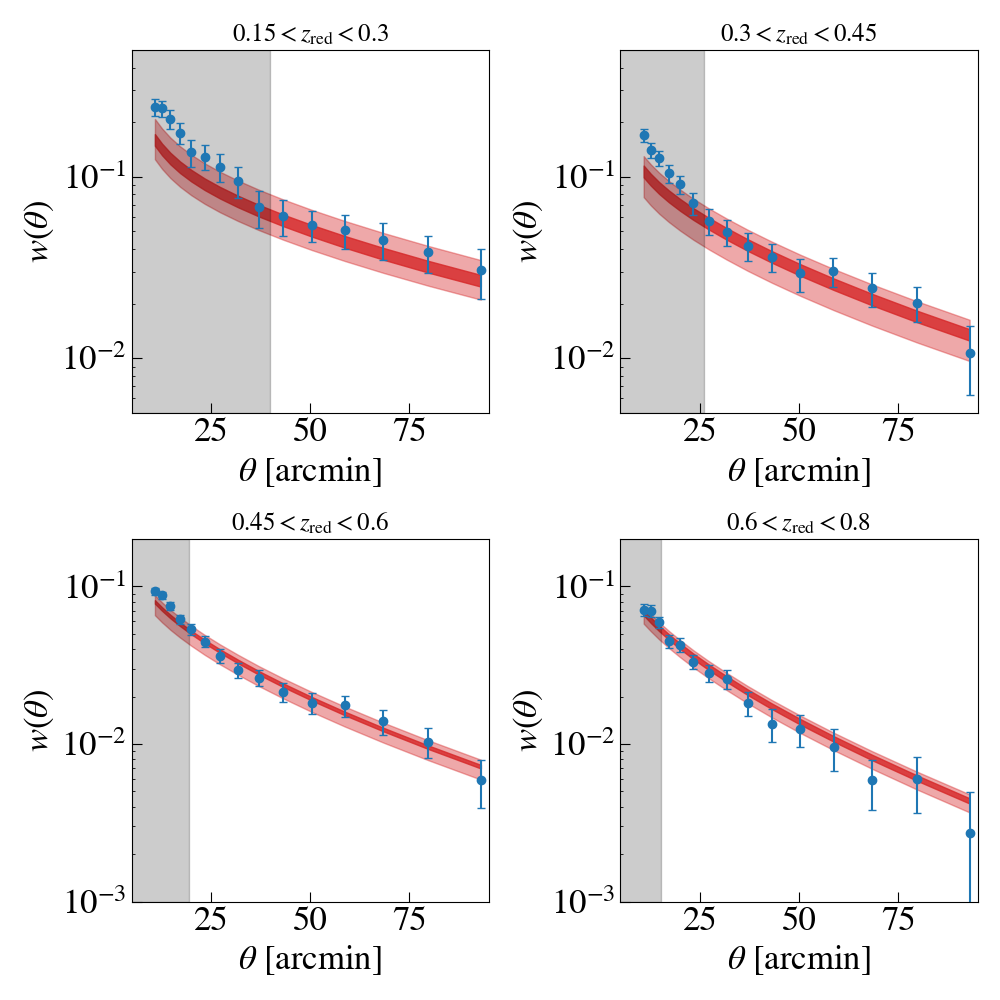}
\caption{ Comparison between the posterior predictions (red shaded) of clustering and the clustering measurements (blue points with error bars) in the four redshift bins. The dark and light shaded regions mark the 68\% and the 95\% confidence intervals. The error bars are derived from the diagonal elements of the blinded covariance matrix of the observations. } 
\label{fig:w_estimate}
\end{figure*}

\subsubsection{Likelihood and posterior sampling}

With the theoretical model, the measurements, and the blinded covariance matrix at hand, now we are ready to constrain the linear galaxy bias and photo-z distribution shift parameters for each redshift bin: $\{b_{g}^{(i)}, \delta z^{(i)}\}_{i=1}^{4}$. 


We assume that the likelihood is a multivariate Gaussian distribution with the mean given by the theoretical prediction (Eq.~\ref{eq:clustering_theory}) and with the inverse covariance given by the blinded inverse covariance matrix (Eq.~\ref{eq:blinded}). The model parameters are constrained by MCMC sampling from the posterior probability distribution $p(\theta | d) \propto p(d|\theta)p(\theta)$ using the $\mathtt{emcee}$ implementation (\citealt{emcee}) of the affine-invariant ensemble Markov Chain Monte Carlo sampling method of~\citet{goodman2010}.

\subsection{Constraints}

\begin{figure}
\includegraphics[width=\columnwidth]{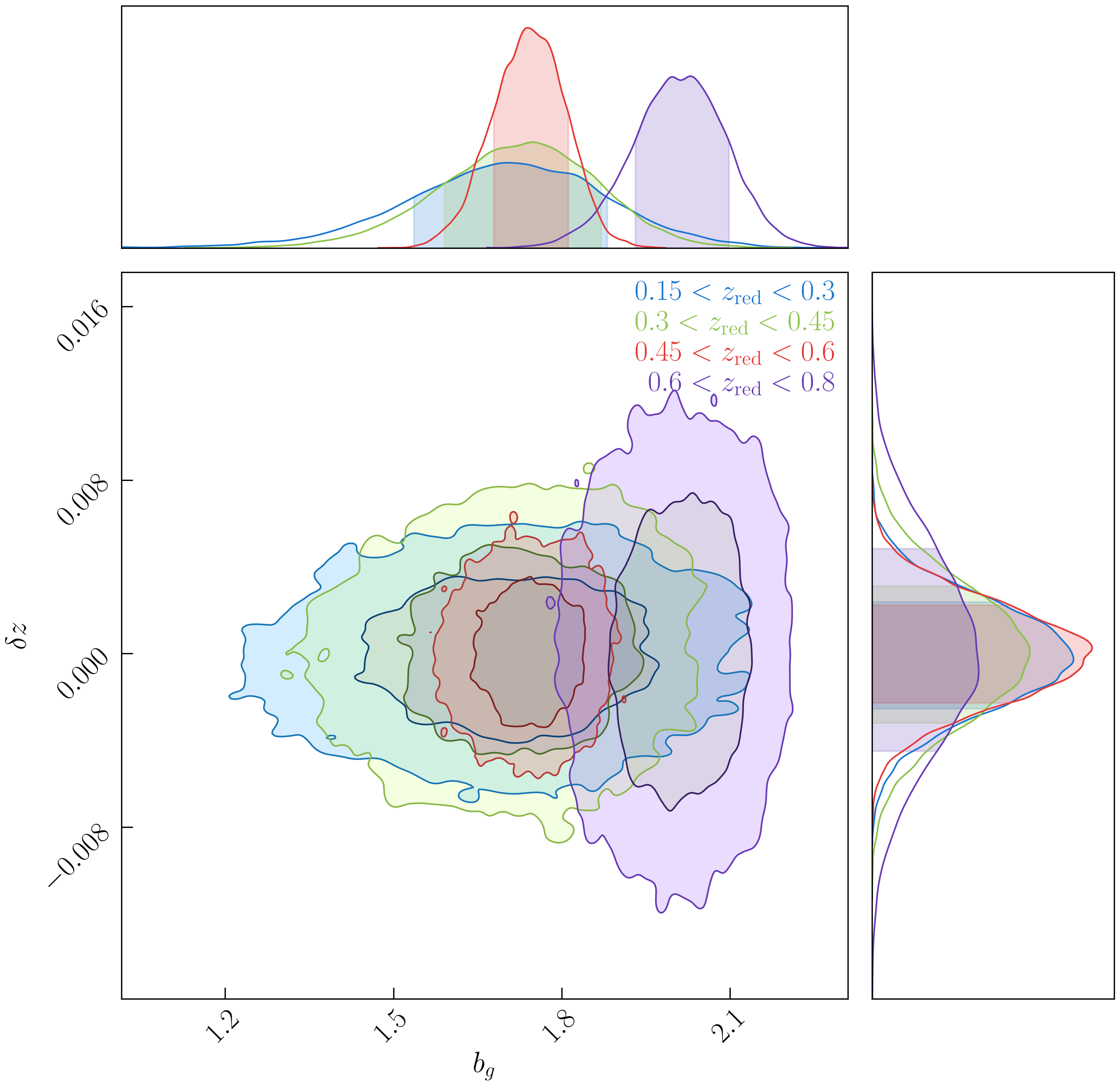}
\caption{ Joint constraints on the linear galaxy bias and photo-z shift parameters shown for the redshift bins constructed with the dense sample (blue contours) and the last redshift bin which is constructed from the luminous sample (red contour).} 
\label{fig:joint_estimate}
\end{figure}

Figure~\ref{fig:w_estimate} presents measurement of the angular correlation function (data points with error bars) together with the 68\% and 95\% posterior predictions of $w_{\theta}$ for all the four redshift bins. Given the uncertainties, the model predictions are consistent with the measurements. 
The 1$\sigma$ and $2\sigma$ levels of the 2D posterior surfaces in the $(b_g, \delta_z)$ parameter space as well as the marginalized distributions over the individual parameters are displayed in Fig.~\ref{fig:joint_estimate}. The correlation between the inferred bias parameter and the photo-z shift parameter appears to be very small. The Spearman correlation\footnote{The Spearman correlation coefficient provides an estimate of the monotonicity in the relation between two parameters without assuming that the distribution of the parameters is Normal (\citealt{zwillinger1999crc}).} between the two parameters is 0.02, 0.06, 0.06, and 0.06 for the four bins in increasing redshift order. 

The marginalized distributions are summarized in Table~\ref{tab:constraints}. We note that the constraints on the photo-z shift parameters are consistent with zero and largely consistent with the adopted priors over these parameters. Furthermore, we find consistency, given the uncertainties, between the constraints on the galaxy bias parameters of the first three bins. Note that the first three bins are constructed from the dense galaxy sample which has an approximately constant comoving density. On the other hand, our constraint on the bias parameter of the last bin is higher than those of the first three bins. This is expected as the last bin is constructed from the luminous sample which consists of brighter galaxies residing in more massive halos.


\begin{table}
	\centering
	\caption{Summary of parameter constraints}
	\label{tab:constraints}
	\begin{tabularx}{0.8\columnwidth}{lcr} 
		\hline
		Redshift bin & $b_g$ & $\delta_z$\\
		\hline
		$0.15<z_{\rm red}<0.3$ & $1.70^{+0.18}_{-0.17}$ & $-0.00^{+0.002}_{-0.002}$\\
		$0.3<z_{\rm red}<0.45$ & $1.72^{+0.14}_{-0.13}$ & 
		$-0.00^{+0.003}_{-0.003}$ \\
        $0.45<z_{\rm red}<0.8$ & $1.74^{+0.06}_{-0.06}$ & $-0.00^{+0.002}_{-0.002}$\\
        $0.6<z_{\rm red}<0.8$ & $2.01^{+0.08}_{-0.08}$ & $0.00^{+0.005}_{-0.004}$\\
		\hline
	\end{tabularx}
	\tablefoot{Model parameter constraints and uncertainties derived from the median and the 68\% confidence intervals of the marginalized posterior distributions.}
\end{table}

In the default setup for studying the large-scale structure, we have relied on four redshift bins, three of which are constructed from the dense sample with bins defined by the redshift edges of [0.15, 0.3, 0.45, 0.6], and one last bin constructed from the luminous sample within the [0.6, 0.8] redshift interval. In order to assess the redshift-dependence of the estimated bias parameters, we define two additional redshift bins for the galaxies in the luminous sample with the redshift edges of [0.2, 0.4, 0.6]. Following the steps we have discussed for our fiducial large-scale structure analysis, we perform redshift estimation, photometric weight assignment, correlation function measurement, and blinding for the two new redshift bins in the luminous sample.  

\begin{figure}
\includegraphics[width=\columnwidth]{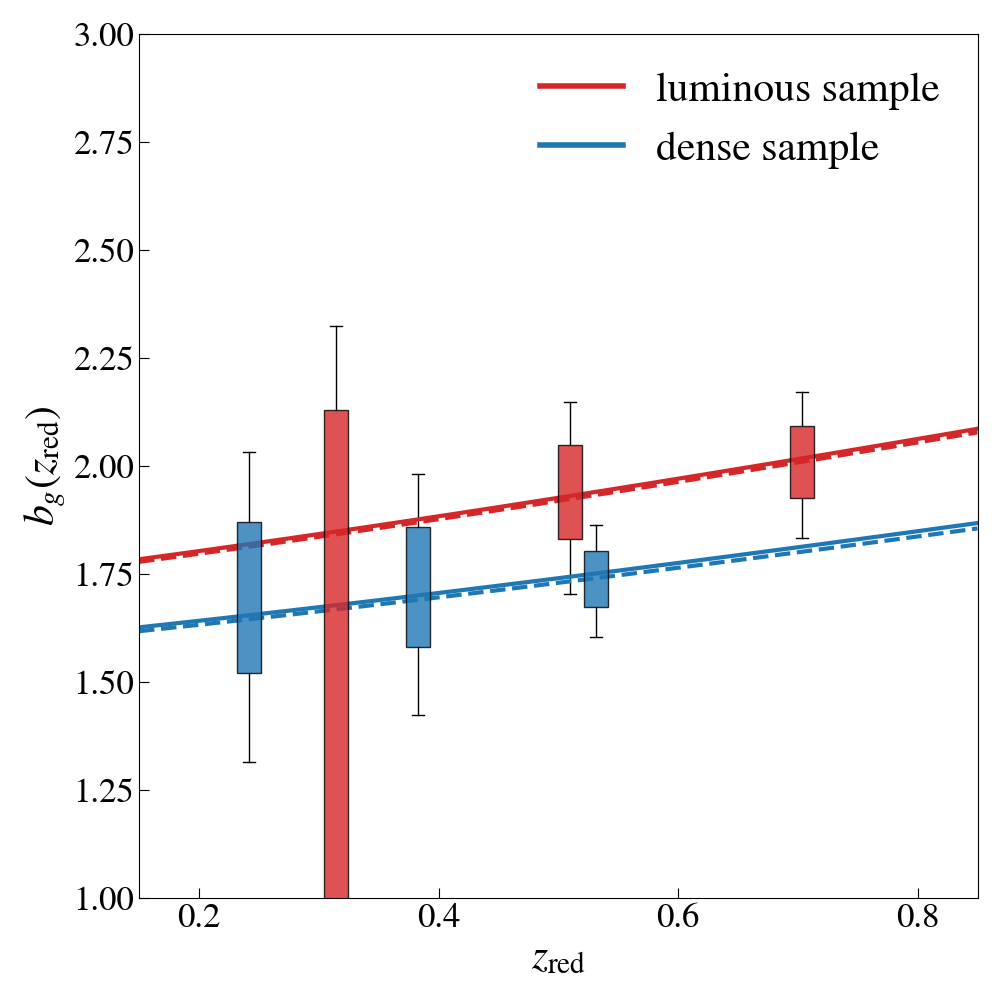}
\caption{ Redshift-dependence of the estimated linear galaxy bias shown for the dense sample (blue) and the luminous sample (red). The boxes mark the 68\% and the 95\% confidence intervals of the marginalized distribution over bias parameters. The solid (dashed) lines show the predictions of the passive evolution model of \citet{Fry1996} given the bias constraints in the three redshift bins (in the last redshift bin) for each galaxy sample.} 
\label{fig:b_estimate}
\end{figure}

The redshift-dependence of the bias is shown in Fig.~\ref{fig:b_estimate}, where the estimated bias parameters of the luminous (dense) sample are shown in red (blue). The boxes mark the 68\% as well as the 95\% confidence intervals in the marginalized bias distributions. We note that within each sample, the bias constraints do not appear to have any strong redshift evolution. We also note that for the dense sample, our linear bias constraints are consistent with the findings of~\citet{brown2008red} that studied the clustering of photometrically selected red galaxies in different ranges of redshift and comoving density. 

We compare our findings with the passive evolution model of galaxy bias (see \citealt{Fry1996, Tegmark1998}) according to which, the evolution of the linear bias follows:
\begin{equation}
b(z) = 1 + \frac{b_{0}-1}{D(z)},
\end{equation}
where $D(z)$ is the linear growth factor normalized to unity at $z=0$, and $b_{0}$ the linear bias at $z=0$.  
The passive evolution model has been tested against the amplitude of the clustering of LRGs in \citet{Rita2012, Guo2013}. In particular, this latter study investigated the evolution of the large-scale bias of red galaxies for various ranges of constant comoving density, absolute magnitude, as well as colour. Following the approach of \citet{Guo2013} we fit the passive evolution model to our estimated biases as a function of redshift in two samples.   

Let us start by describing the bias evolution of the luminous sample. By fitting the passive evolution model of \citet{Fry1996} to the bias constraints in the luminous sample we find that this model provides a good picture of the evolution of galaxy bias yielding a $\chi^2/\mathrm{dof}$ of 0.04/2 (red solid line in Fig.~\ref{fig:b_estimate}). When constraining the passive evolution model with the bias estimate of the highest redshift bin ($0.6<z_{\rm red}<0.8$), we find that the expected bias parameters for the first and the second bins are consistent with the estimated parameters within 1$\sigma$ levels (red dashed line in Fig.~\ref{fig:b_estimate}). 

Turning our attention to the dense sample, we find that the passive evolution model still provides a consistent picture of bias evolution with a $\chi^2/\mathrm{dof}$ of 0.17/2 (blue solid line in Fig.~\ref{fig:b_estimate}). Once we condition the passive evolution model on the bias constraint of the highest redshift bin ($0.45<z_{\rm red}<0.6$), we find that the expected bias values for the first and the second redshift bins are in perfect agreement with the inferred bias parameters (blue dashed line in Fig.~\ref{fig:b_estimate}). Overall, we find that the redshift evolution of the bias of our LRG samples is consistent with the passive evolution model. The small $\chi^2/\mathrm{dof}$ values can also be attributed to the large error bars on the estimated biases of the two samples.


\section{Summary}\label{sec:summary} 

In this work we have introduced the selection and clustering measurements of the red-sequence galaxies in the fourth data release of the Kilo-Degree Survey. The data-driven colour-redshift relation of these galaxies allows us to obtain precise and accurate estimates of their redshifts. We construct two samples, a bright one and a dense one, each with approximately constant comoving density.

We find that the near-infrared magnitudes derived from the VIKING imaging of the selected galaxies allow us to assess the purity of the sample. This purity assessment is done by comparing the colour distribution of the red-sequence candidates and that of high confidence stars in the fourth data release. The outcome of this procedure is the removal of $\sim40\%$ of the candidates in the dense sample with $z_{\rm red}>0.6$ and $\sim5\%$ of the the candidates in the luminous sample in the same redshift range.

After taking into account the purity and completeness of the samples, we construct four redshift bins for our large-scale structure analysis with the first three bins based on the dense sample and the last bin based on the bright sample. 
In order to estimate the redshift distributions as well as the uncertainty over the mean redshift of the distributions, we rely on four spectroscopic redshift surveys. Of these redshift surveys, three have overlap with the fourth data release, while GAMA-G10 only covers one of the KiDS calibration fields in the COSMOS region. In each redshift bin, the individual redshift distributions of galaxies are well described by a Student t-distribution, parameters of which are estimated with the overlapping spectroscopic data sets. 

In order to account for the impact of data quality, we extend the works of \citet{bautista2018sdss} and \citet{icaza2020clustering} to allow for a more flexible relation between the systematic-induced variations of observed galaxy densities and the survey properties, while making use of heavy regularization to avoid overfitting. In comparison to \citet{ross2012clustering, crocce2019dark}, our adopted framework for removing the impact of survey properties does not make any assumption regarding the lack of correlation between the survey properties. Having validated our method for removing the effect of imaging systematics on the observed density variations, we apply the derived photometric weights to the measurement of the red-sequence galaxy clustering.

In order to avoid confirmation bias in our theoretical interpretation of the clustering measurements, we adopt a blinding method introduced by \citet{sellentin2019} in which the estimated inverse covariance matrix of the clustering measurements is perturbed. This perturbation manifests itself in shifting the posterior probability distributions over model parameters. 

We find that the estimated bias parameters of the galaxies in the $L>0.5L_{\rm pivot}(z)$ sample are lower than those of $L>L_{\rm pivot}(z)$ sample, consistent with the expectation that brighter galaxies reside in higher mass halos. The constraints on the photo-z shift parameters are consistent with zero and largely consistent with the adopted priors over these parameters. By comparing the redshift evolution of our bias constraints with the passive evolution model, we find that the bias evolution of galaxies in both dense and luminous samples is consistent with the expectations of the model.

We will utilize the large-scale analysis of this study in a 3$\times$2pt analysis for constraining the cosmological parameters with combination of the positions of red-sequence galaxies and cosmic shear signal of the background sources in the fourth data release of the Kilo-Degree Survey. The galaxy sample constructed in this work is also being used for constraining the intrinsic alignment of galaxies (Fortuna et al. in prep.) and constraining cosmological parameters with density-split statistics (Burger et al. in prep.).

\section*{Acknowledgements}

MV, HHo and MCF acknowledge support from Vici grant 639.043.512 from the Netherlands
Organization of Scientific Research (NWO).
MB is supported by the Polish Ministry of Science and Higher Education through grant DIR/WK/2018/12, and by the Polish National Science Center through grant no. 2018/30/E/ST9/00698. 
HJ acknowledges support from a UK Science and Technology Facilities Council (STFC) Studentship. 
AHW is supported by an European Research Council Consolidator Grant (No. 770935). 
MA acknowledges support from the European Research Council under grant number 647112. 
BG acknowledges support from the European Research Council under grant number 647112 and from the Royal Society through an Enhancement Award (RGF/EA/181006). 
CH acknowledges support from the European Research Council under grant number 647112, and support from the Max Planck Society and the Alexander von Humboldt Foundation in the framework of the Max Planck-Humboldt Research Award endowed by the Federal Ministry of Education and Research. 
H. Hildebrandt is supported by a Heisenberg grant of the Deutsche Forschungsgemeinschaft (Hi 1495/5-1) as well as an ERC Consolidator Grant (No. 770935). 
SJ acknowledges support from the Beecroft Trust and ERC 693024.

GAMA is a joint European-Australasian project based
around a spectroscopic campaign using the Anglo Australian
Telescope. The GAMA input catalogue is based on data
taken from the Sloan Digital Sky Survey and the UKIRT
Infrared Deep Sky Survey. Complementary imaging of the
GAMA regions is being obtained by a number of independent survey programs including GALEX MIS, VST
KiDS, VISTA VIKING, WISE, Herschel-ATLAS, GMRT
and ASKAP providing UV to radio coverage. GAMA is
funded by the STFC (UK), the ARC (Australia), the AAO,
and the participating institutions. The GAMA website is
\hyperlink{www.gama-survey.org}{www.gama-survey.org}.

Funding for SDSS-III was provided by the Alfred P.
Sloan Foundation, the Participating Institutions, the National Science Foundation, and the U.S. Department of
Energy Office of Science. The SDSS-III website is \hyperlink{http:
//www.sdss3.org/}{http:
//www.sdss3.org/}. SDSS-III is managed by the Astrophysical Research Consortium for the Participating Institutions
of the SDSS-III Collaboration including the University of
Arizona, the Brazilian Participation Group, Brookhaven
National Laboratory, Carnegie Mellon University, University of Florida, the French Participation Group, the German Participation Group, Harvard University, the Instituto de Astrofisica de Canarias, the Michigan State/NotreDame/JINA Participation Group, Johns Hopkins University, Lawrence Berkeley National Laboratory, Max Planck
Institute for Astrophysics, Max Planck Institute for Extraterrestrial Physics, New Mexico State University, New
York University, Ohio State University, Pennsylvania State
University, University of Portsmouth, Princeton University,
the Spanish Participation Group, University of Tokyo, University of Utah, Vanderbilt University, University of Virginia, University of Washington, and Yale University.

This work has made use of \hyperlink{www.python.org}{python},
including the packages \hyperlink{www.numpy.org}{numpy}, \hyperlink{www.scipy.org}{scipy}, \hyperlink{https://pandas.pydata.org/}{pandas}, and \hyperlink{https://scikit-learn.org/}{scikit-learn}. 
Plots have been produced with \hyperlink{matplotlib.org}{matplotlib} and \hyperlink{https://seaborn.pydata.org/}{seaborn}. We have used the ChainConsumer package (\citealt{Hinton2016}) to generate Fig.~\ref{fig:joint_estimate} of the paper.
This work has made use of \hyperlink{cosmohub.pic.es}{CosmoHub}.
CosmoHub has been developed by the Port dInformacio Cientifica (PIC), maintained through a collaboration of the Institut de Fisica d Altes Energies (IFAE) and the Centro de Investigaciones Energeticas, Medioambientales y Tecnologicas (CIEMAT) and the Institute of Space Sciences (CSIC \& IEEC), and was partially funded by the "Plan Estatal de Investigacion Cientifica y Tecnica y de Innovacion" program of the Spanish government.

\textit{Author contributions}: All authors contributed to the development and writing of this paper. The authorship list is given in three groups: the lead authors (MV,HHo, MB, MCF) followed by two alphabetical groups. The first alphabetical group includes those who are key contributors to both the scientific analysis and the data products. The second group covers those who have either made a significant contribution to the data products, or to the scientific analysis.

\bibliographystyle{mnras}
\bibliography{lrg_kids.bib}



\appendix

\section{Comparison between the blinded and unblinded results}\label{A:blind}

\begin{figure}
\begin{center}
\includegraphics[width=0.9\textwidth, height = 0.4\textwidth]{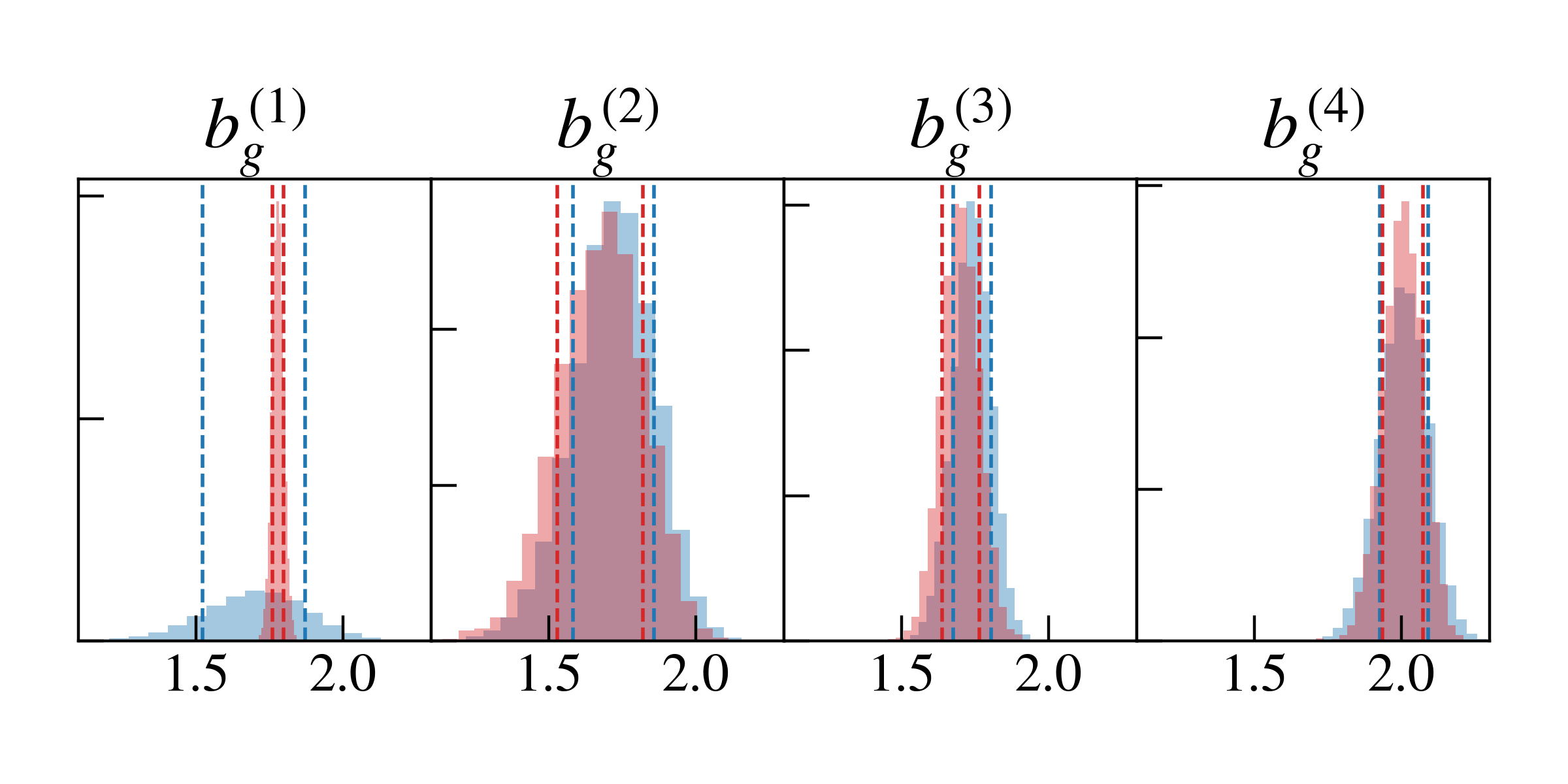}
\caption{Illustration of the difference between the unblinded (blue histogram) and the blinded (red histogram) marginal posteriors over the bias parameters, along with the 68\% confidence intervals (dashed lines).}
\label{fig:blind_vs_unblind_post}
\end{center}
\end{figure}

Figure~\ref{fig:blind_vs_unblind_post} demonstrates the difference between the marginalized posterior distributions of the galaxy bias parameters in our sample given the blind covariance matrix (shown in red) and the unblind covariance (shown in blue). Generally, the blinded posteriors are slightly shifted with respect to the unblinded ones except for the posterior probability of $b_{g}^{(1)}$. The narrow width of the blinded constraint on $b_{g}^{(1)}$ is due to the fact that for the first redshift bin, the last diagonal element of the matrix $B$ given by Eq.~\ref{eq:bii} is large resulting in the collapse of the last diagonal element of the blinded covariance matrix. Consequently, this gives rise to a narrower constraint on $b_{g}^{(1)}$ given the blinded covariance matrix. In order for this blinding method to work as expected, the diagonal elements of the matrix $B$ need to be close to unity which is not the case in the first redshift bin. In principle, this problem could have been avoided by imposing further constraints such as the invariance of the width of the posterior after blinding using the procedure described in Sect. 4.3 of~\citet{sellentin2019}. In our blinding procedure, we had set the parameter $b_{\rm fid}$ ($\tilde{b}$) to 1.80 (1.78) for first three redshift bin and 2.1 (2.08) for the last redshift bin. That is, if the true underlying bias parameter of a redshift bin was $b_{\rm fid}$, the blinding scheme would result in a -0.02 shift in the position of the peak of the posterior probability.


\label{lastpage}

\end{document}